\documentclass[iop]{emulateapj}

\usepackage{graphicx}
\usepackage{epstopdf}
\newcommand{\water}{H$_2$O}
\newcommand{\ozone}{O$_3$}
\newcommand{\ammonia}{NH$_3$}
\newcommand{\methanol}{CH$_3$OH}
\newcommand{\invcm}{cm$^{-1}$}

\newcommand{\av}{$A_V$}
\newcommand{\ejk}{$E(J-K)$}
\newcommand{\hii}{\ion{H}{2}}

\slugcomment{Accepted by The Astrophysical Journal, scheduled to appear in V730, March 20, 2011.}

\shorttitle{Ice in IC~5146}
\shortauthors{Chiar et al.}

\begin{document}


\title{Ices in the Quiescent IC 5146 Dense Cloud}


\author{J. E. Chiar\altaffilmark{1},
Y.J.\ Pendleton\altaffilmark{2}, 
L.J.\ Allamandola\altaffilmark{2}, 
A.C.A.\ Boogert\altaffilmark{3},
K.\ Ennico\altaffilmark{2}, 
T.P.\ Greene\altaffilmark{2}, 
T.R.\ Geballe\altaffilmark{4},
J.V.\ Keane\altaffilmark{5},
C.J.\ Lada\altaffilmark{6},
R.E.\ Mason\altaffilmark{4},
T.L.\ Roellig\altaffilmark{2}, 
S.A.\ Sandford\altaffilmark{2},
A.G.G.M.\ Tielens\altaffilmark{7},
M.W.\ Werner\altaffilmark{8},
D.C.B.\ Whittet\altaffilmark{9}
L.\ Decin\altaffilmark{10},
K.\ Eriksson\altaffilmark{11},
}

\altaffiltext{1}{SETI Institute, Carl Sagan Center, 189 Bernardo Avenue, Mountain View, CA 94035, USA {jchiar@seti.org}}
\altaffiltext{2}{NASA Ames Research Center, MS 245-6, Moffett Field, CA, 94035, USA}
\altaffiltext{3}{IPAC, NASA Herschel Science Center, Mail Code 100-22, California Institute of Technology, Pasadena, CA 91125, USA}
\altaffiltext{4}{Gemini Observatory, Northern Operations Center, 670 N. A'ohoku Place, Hilo, HI 96720, USA}
\altaffiltext{5}{Institute for Astronomy, 2680 Woodlawn Drive, Honolulu, Hawaii 96822, USA}
\altaffiltext{6}{Harvard-Smithsonian Center for Astrophysics, 60 Garden Street, Cambridge, MA 02138, USA}
\altaffiltext{7}{Leiden Observatory, P.O. Box 9513, 2300 RA Leiden, The Netherlands}
\altaffiltext{8}{Jet Propulsion Laboratory, California Institute of Technology, 4800 Oak Grove Drive, Pasadena, CA 91109, USA}
\altaffiltext{9}{New York Center for Astrobiology, Department of Physics, Applied Physics \& Astronomy, Rensselaer Polytechnic Institute, Troy, N.Y., 12180, USA}
\altaffiltext{10}{Instituut voor Sterrenkunde, Katholieke Universiteit Leuven, Celestijnenlaan 200D, 3001 Leuven, Belgium}
\altaffiltext{11}{Department of Physics \& Astronomy, Uppsala University, Box 515, 751 20 Uppsala, Sweden}


\begin{abstract}
This paper presents spectra in the 2 to 20 \micron\ range of quiescent cloud material located in the IC 5146 cloud complex.  The spectra were obtained with NASA's Infrared Telescope Facility (IRTF) SpeX instrument and the Spitzer Space Telescope's Infrared Spectrometer.   We use these spectra to investigate dust and ice absorption features in pristine regions of the cloud that are unaltered by embedded stars.   We find that the \water-ice threshold extinction is $4.03\pm0.05$ mag.  Once foreground extinction is taken into account, however, the threshold drops to 3.2 mag, equivalent to that found for the Taurus dark cloud, generally assumed to be the touchstone quiescent cloud against which all other dense cloud and embedded young stellar object observations are compared.  Substructure in the trough of the silicate band for two sources is attributed to \methanol\ and \ammonia\ in the ices, present at the $\sim2$\% and $\sim5$\% levels, respectively, relative to \water-ice.  The correlation of the silicate feature with the $E(J-K)$ color excess is found to follow a much shallower slope relative to lines of sight that probe diffuse clouds, supporting the previous results by Chiar et al.\ (2007).

\end{abstract}





\section{Introduction}

Ices are abundant in preplanetary material and their presence is likely to represent a vital resource for the origin of planetary life \nocite{mottl2007} (e.g., Mottl et al.\ 2007). Ices are widely observed in the dense molecular clouds that give birth to new stars \nocite{gibb2004} (e.g., Gibb et al.\ 2004); they are ubiquitous in comets \nocite{whipple1950,schulz2006} (e.g., Whipple 1950; Schulz et al.\ 2006), and have recently been detected on asteroidal bodies as well \nocite{campins2010,rivkin2010} (Campins et al.\ 2010; Rivkin \& Emery 2010). To determine the evolution of such materials, from icy coatings on submicron-sized interstellar grains to planetesimals in protoplanetary disks, is a major goal in the astronomical search for origins. A crucial first step is the formation of ice mixtures composed primarily of H$_2$O, CO and CO$_2$ in molecular clouds. The observed abundances of these molecules in the solid phase can be explained by simple grain surface reactions such as ${\rm O + H \rightarrow OH}$, ${\rm OH + H \rightarrow H_2O}$ and ${\rm CO + OH \rightarrow CO_2 + H}$, in combination with direct freeze-out of CO from the gas phase.

The physical conditions under which these reactions occur may vary from cloud to cloud, as a function of factors such as the cloud mass, temperature, and the intensity of the local interstellar radiation field. Ices are quite widely observed in the spectra of young stellar objects (YSOs), but the study of ices in pristine {\it prestellar\/} dark clouds requires observations of background field stars in lines of sight that do not intercept circumstellar material around YSOs. Relatively few such observations are currently available. The dark cloud in Taurus is by far the best studied to date \nocite{whittet1988,whittet2007,chiar1995,bergin2005,shenoy2008} (e.g., Whittet et al.\ 1988, 2007; Chiar et al.\ 1995; Bergin et al\ 2005; Shenoy et al.\ 2008) and is often assumed to be prototypical, but this remains to be confirmed.  Observations of ices toward a significant sample of background stars are available for only two other molecular clouds: Serpens and $\rho$~Oph \nocite{eiroa1989,tanaka1990,chiar1994,knez2005} (Eiroa \& Hodapp 1989; Tanaka et al\ 1990; Chiar et al.\ 1994; Knez  et al.\ 2005).  Boogert et al.\ (2011) \nocite{boogert2011} present a study of ices and extinction toward sixteen isolated cores as probed by background stars and find the threshold extinction to be in line with that found in Taurus. The observations show similarities but also significant differences in ice properties and the conditions needed for ice to form. These differences may be parameterized in terms of the ice ``extinction threshold," i.e., the minimum observed continuum extinction necessary for the detection of the ices \nocite{whittet2003} (e.g., Whittet 2003), and in the relative abundances of H$_2$O, CO and CO$_2$ \nocite{whittet2007,whittet2009} (Whittet et al.\ 2007, 2009). To establish a reliable benchmark for studies of ice evolution in regions of active star formation it is vital to understand the initial conditions in quiescent clouds and the extent to which they vary from cloud to cloud. 

With this goal in mind, we present new observations of ices in the IC 5146 dark cloud complex. IC 5146 is nearly 3\arcdeg\ in extent, located in the direction of Cygnus \nocite{elias1978,dobashi1992} (Elias 1978; Dobashi et al.\ 1992).   The IC 5146 complex consists of several dark clouds as well as the well-known \hii\ region, the Cocoon Nebula (Dobashi et al.\ 1992).  The dark cloud called the Northern Streamer,  a region of high extinction compared to the rest of the cloud, is located to the west of the Cocoon \nocite{lada1999,harvey2008,cambresy1999}(Cambr{\'e}sy 1999; Lada et al. 1999; Harvey et al.\ 2008).   It contains several protostellar clumps, some of which are actively forming stars (Dobashi et al.\ 1992).  These distinct regions are thought to be co-distant at 950~pc \nocite{harvey2008} (Harvey et al.\ 2008).  Star-formation in the \hii\ region is loosely clustered and more efficient than that in the Northwest Streamer where most of the YSOs have formed in isolation. The field stars presented in this paper probe this relatively quiescent Northwest Streamer region.  In \S\ref{sec:observations} we describe the complete data set for the IC 5146 cloud including the Spitzer (IRS, IRAC, MIPS), 2MASS \nocite{skrutskie2006} (Skrutskie et al.\ 2006), and IRTF-SpeX observations.  We used our IRTF-SpeX spectra to classify the spectral types of the background stars (\S\ref{sec:sptype}), thus allowing us to fit the continuum flux using a reddened photospheric model appropriate for the spectral type of the star (\S\ref{sec:continua}).  The resulting optical depth profiles and ice abundances are discussed in \S\ref{sec:iceprofiles}.  The relationship between the near-infrared color excess and the silicate optical depth is discussed in \S\ref{sec:extinction}. Finally, we summarize our results in \S\ref{sec:final}.

%%% TABLE 1
\begin{deluxetable*}{cccccc}
\tabletypesize{\scriptsize}
\tablecolumns{6}
%\rotate
\tablewidth{0pc}
\tablecaption{Field Stars located behind IC 5146\label{table:sources}}
\tablehead{
\colhead{2MASS ID} & \colhead{Spitzer AOR label} & 
\colhead{Spectral Type\tablenotemark{a}} & 
\colhead{$E(J-K)$\tablenotemark{b}} & 
\colhead{$A_V$ from model match} &
\colhead{AOR Key}}
\startdata
21472204+4734410 & Q21-1 & K2 III Fe-0.5 & $4.5^{+0.07}_{-0.07}$ & 23.9 & 14136064 \\[1.6ex]
21463943+4733014 & Q21-2 & G8 III Fe-1 & $2.3^{+0.05}_{-0.00}$ & 12.2 & 10750720  \\[1.6ex]
21475842+4737164 & Q21-3 & [K2 III] & 2.1 & 11.1 & 10750720 \\[1.6ex]
21450774+4731151 & Q21-4 & K6 III CN 0.5 & $1.6^{+0.08}_{-0.06}$ & 8.5 & 10750720 \\[1.6ex]
21444787+4732574 & Q21-5 & K2 III Fe-0.5 & $1.5^{+0.07}_{-0.07}$ & 8.0 &  14136064 \\[1.6ex]
21461164+4734542 & Q21-6 & G8.5 IIIa Fe-0.5 & $3.9^{+0.05}_{-0.00}$ & 20.7 & 14136064 \\[1.6ex]
21443293+4734569 & Q22-1 & K0 III & $3.3^{+0.04}_{-0.05}$ & 17.49 & 14136064 \\[1.6ex]
21473989+4735485 & Q22-3 & [K1 III] & 0.7 & 3.7 & 10750976 \\[1.6ex]
21473509+4737164 & Q23-1 & K1 IIIb CN1.5 Ca1 & $1.4^{+0.07}_{-0.04}$ & 7.4 & 14136064 \\[1.6ex]
21472220+4738045 & Q23-2 & [K5 III] & 1.9 & 10.1 & 14136064 \\
\enddata
%% Text for table notes should follow after the \enddata but before
%% the \end{deluxetable}. Make sure there is at least one \tablenotemark
%% in the table for each \tablenotetext.
\tablenotetext{a}{Spectral type based on match to standard star spectra in IRTF Spectral Library, with reddening applied. Square brackets indicate no IRTF data were available to classify the star and spectral type is based on fit of reddened Decin models to photometry and Spitzer IRS spectra.}
\tablenotetext{b}{Intrinsic colors from Tokunaga 2000}
\end{deluxetable*}

\section{Observations and Data Reduction}
 \label{sec:observations}

We used the Infrared Spectrometer \nocite{houck2004} (Houck et al. 2004) on the Spitzer Space Telescope \nocite{werner2004} (Werner et al. 2004) to obtain data in the 5 to 20 \micron\ region of 10 sources behind the IC 5146 dark cloud, with visual extinction ($A_V$) in the range from 4 to 27 mag.  Source names, 2MASS IDs, spectral class (see \S\ref{sec:sptype}), extinction information, and AOR keys are listed in Table~\ref{table:sources}.  The IRS pipeline version used was 15.3. The low resolution ÔShort-LowÕ and ÔLong-LowÕ modules (SL and LL; $R\sim60-120$) were reduced in a way that is customary for ground-based spectra, following the method described in \nocite{boogert2008} Boogert et al. (2008).  For the IC 5146 cloud, the background is relatively uniform across the array, so background subtraction was carried out by using the nodded pairs.  A fixed width extraction was performed and the 1-dimensional spectra were then averaged. Subsequently the IC 5146 spectra were divided by the ratioed spectra of the standard stars HR 2194 (A0 V; PID 1417; AOR keys 0013024512 and 0013024768 for SL1 and SL2 respectively) and HR 6606 (G9 III; PID 1421; AOR key 0013731840 for LL) reduced in the same way in order to correct for wavelength-dependent slit losses.  The standard stars were ratioed with the model spectra from Decin et al. (2007) appropriate for their spectral type.

For seven of these sources, we also obtained 1.9 to 4.0 \micron\ data with the SpeX instrument \nocite{rayner2003} (Rayner et al.\ 2003) on NASA's Infrared Telescope Facility (IRTF).  The spectra were obtained using a 0.8\arcsec\ slit with the cross-dispersed LXD mode of the SpeX instrument resulting in $R\sim937$. The spectra were reduced using the SpeXtool software package \nocite{vacca2003spextool,cushing2004spextool} (Vacca et al.\ 2003; Cushing et al.\ 2004). These data were used to classify the spectral types of the stars and measure the 3.0 \micron\ H$_2$O ice absorption feature.  For the most heavily obscured source, Q21-1 we also obtained an M-band spectrum in order to measure the CO-ice profile at 4.67 \micron.

\begin{figure}
\figurenum{1}
\includegraphics[angle=0,scale=0.45]{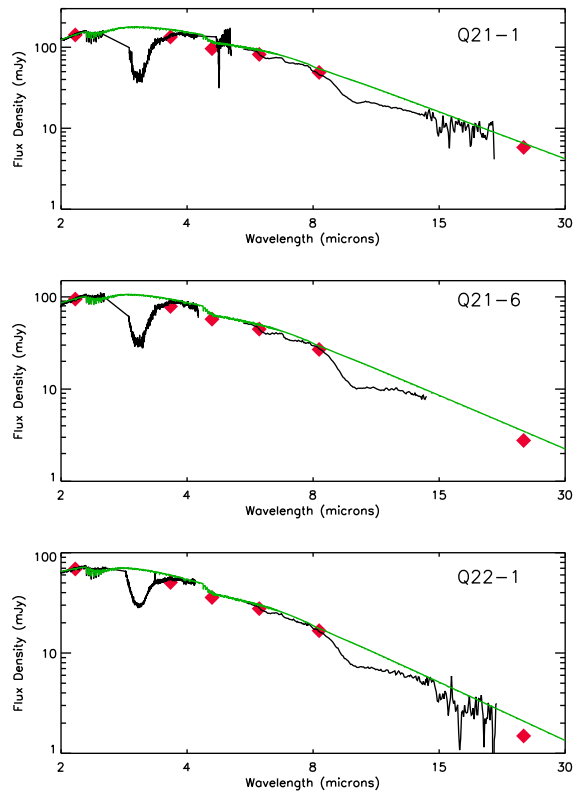}
\caption{Spectral energy distributions for the IC 5146 field stars. 2MASS and IRAC photometry (diamonds), SpeX and IRS spectra (black lines) and the reddened photospheric model  spectrum (green line) are shown.  The spectra are plotted in order of decreasing visual extinction: Q21-1 (\av=23.9 mag), Q21-6 (\av=20.7 mag), Q22-1 (\av=17.49 mag). }
\label{fig:continua}
\end{figure}

In order to flux calibrate the Spitzer-IRS and IRTF-SpeX spectra, we used available photometry from 2MASS and IRAC \nocite{harvey2008} (Harvey et al. 2008).  The convolved flux at the effective wavelength of the photometric filter was calculated using published responsivity curves for 2MASS \nocite{cohen2003,reach2005}(Cohen et al.\ 2003) and  IRAC (Reach et al.\ 2005).  Following the calculation of the effective flux over a particular photometric passband, a normalization factor was calculated individually for the SpeX and IRS spectra. SpeX spectra were normalized to the 2MASS $K_S$ band, and IRS spectra were normalized to either the IRAC3 or IRAC4 band.  The flux calibrated spectra along with the photometry are shown in Fig.~\ref{fig:continua}.

%% FIGURE 1:  spex and irs spectra and continuum fits

\begin{figure}
\figurenum{1}
\includegraphics[angle=0,scale=0.45]{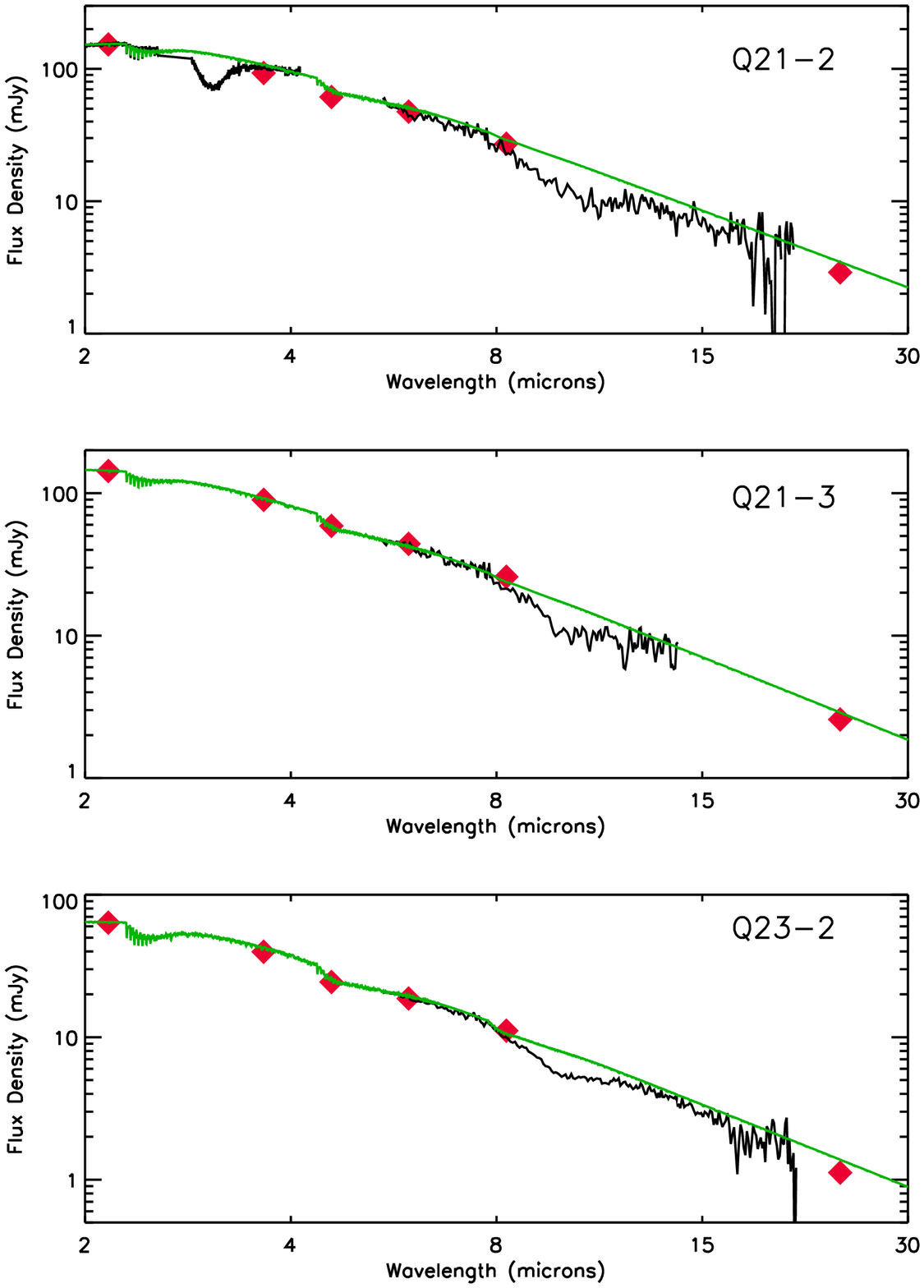}
\caption{Q21-2  (\av=12.2 mag), Q21-3  (\av=11.1 mag), Q23-2  (\av=10.1 mag). }
\end{figure}

\begin{figure}
\figurenum{1}
\includegraphics[angle=0,scale=0.45]{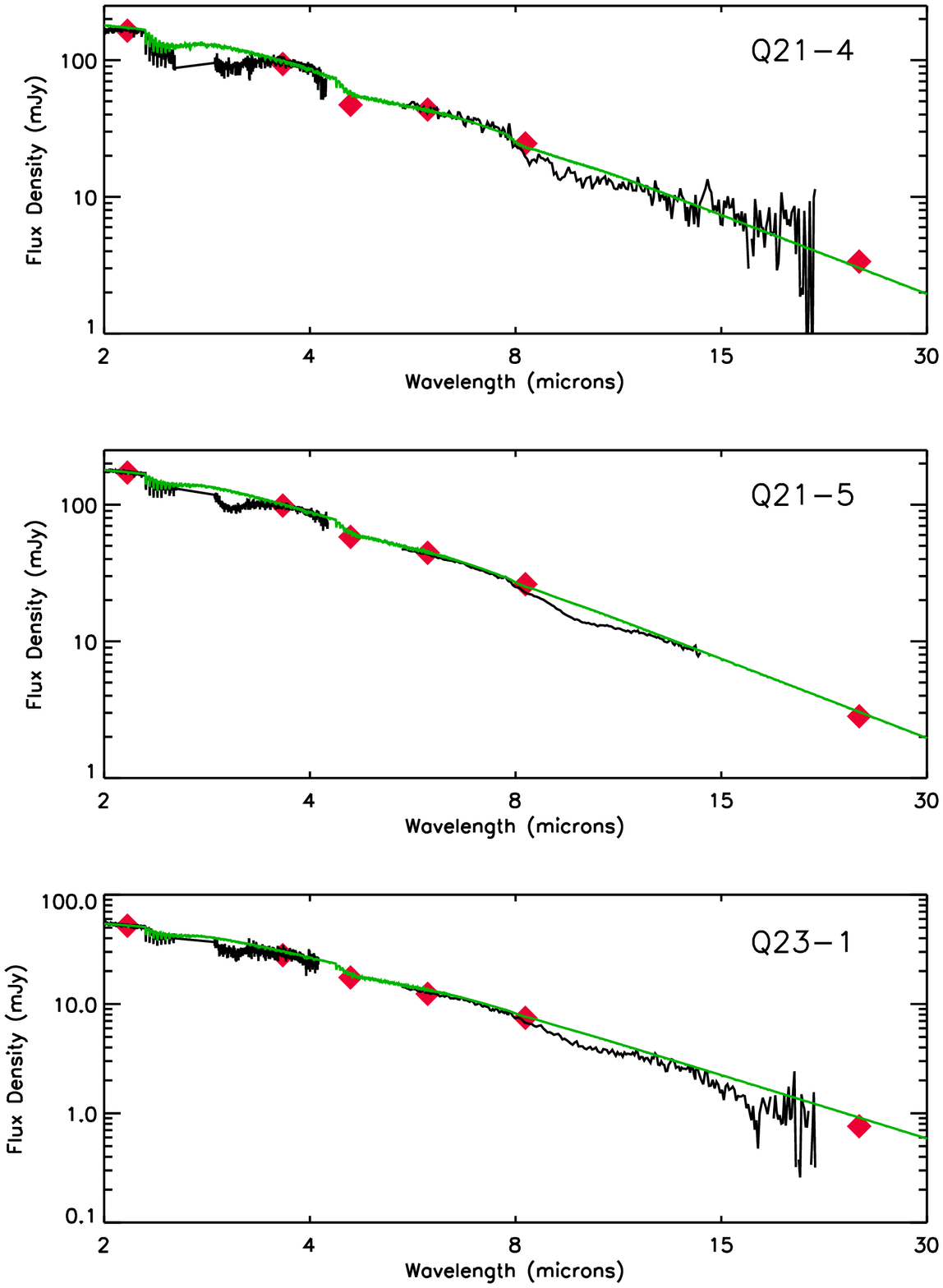}
\caption{Q21-4 (\av=8.5 mag), Q21-5 (\av=8.0 mag), Q23-1 (\av=7.4 mag).}
\end{figure}

\begin{figure}
\figurenum{1}
\includegraphics[angle=0,scale=0.45,trim = 0mm 160mm 0mm 0mm, clip]{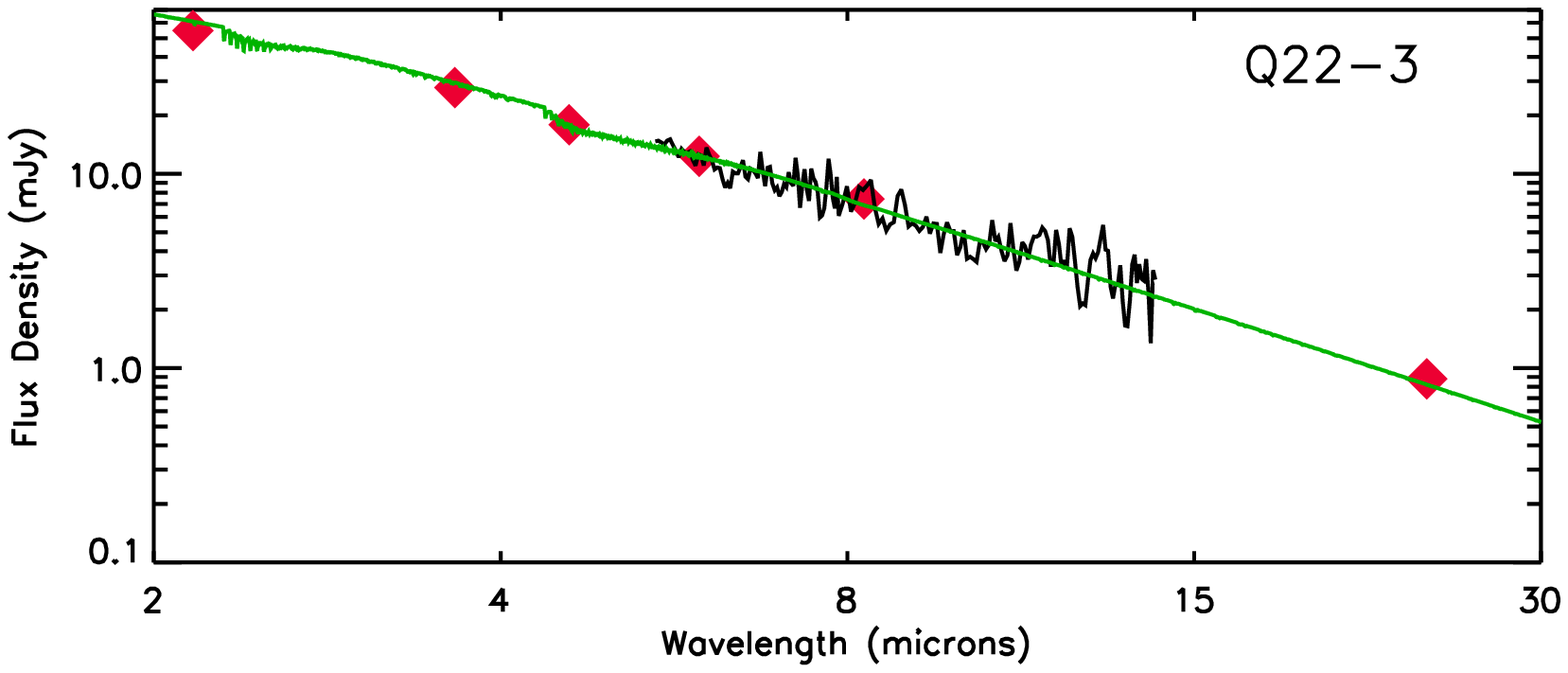}
\caption{Q22-3  (\av=3.7 mag)}
\end{figure}

\subsection{Spectral Classification}
\label{sec:sptype}

\begin{figure}
\figurenum{2}
\begin{center}
\includegraphics[angle=0,scale=0.4]{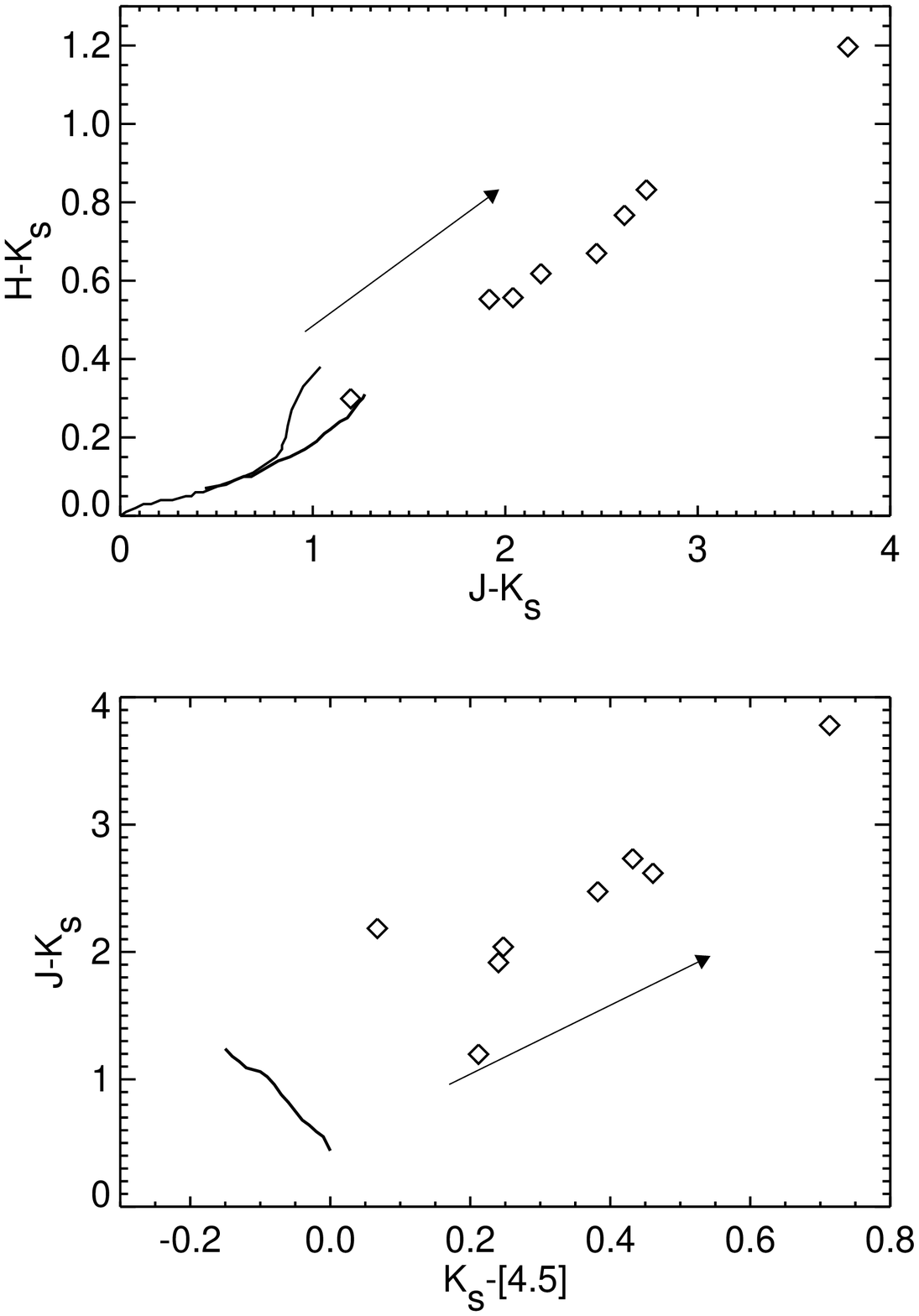}
\caption{[top] $H-K_s$ vs. $J-K$  and [bottom] and $J-K$ vs. $K_s-[4.5]$ for the IC 5146 field stars (diamonds).  In the top panel, smooth lines denote the dwarf (upper curve) and giant (lower curve) branches of normal unreddened stars.  In the bottom panel, the curve for normal giant stars is shown.  The reddening vectors are equivalent to $A_V\sim5$ mag.}
\label{fig:colorcolor}
\end{center}
\end{figure}

In order to confirm the background nature of our sources, we constructed color-color diagrams which separate out reddened field stars from embedded objects and stars with circumstellar material.   Figure \ref{fig:colorcolor} shows the placement of the IC 5146 sources with respect to normal giant and dwarf stars \nocite{bessell1988} (Bessell \& Brett 1988)\footnote{We chose not to transform these intrinsic colors to the 2MASS system because the transformation artificially displaces the ``origin'' of the dwarf branch away from $J-K_s = H-K_s = 0$.}, along with a reddening vector.     Reddening curves are deduced using the extinction law determined by \nocite{indebetouw2005} Indebetouw et al.\ (2005) and are equivalent to $A_V\sim5$ mag.  It is apparent from these diagrams that all the IC 5146 program stars are field giants reddened by dense cloud dust that displaces them from the red-giant branch.  Young stellar objects and objects with circumstellar dust would fall near the bottom right in the $J-K_s$ vs. $[4.5] - K_s$ diagram.

We used the 2.2 to 2.5 \micron\ region to determine the spectral type of the stars for which we have IRTF-SpeX spectra (Fig.~\ref{fig:spectralclass}).  In this spectral region, the strength of the CO lines is a strong indicator of the spectral type (and luminosity class) of late-type stars.  The $^{12}$CO first overtone, spanning the 2.29--2.50 \micron\ region, dominates the K-band spectra of the giant stars \nocite{wallace1996,wallace1997,heras2002} (Wallace \& Hinkle  1996, 1997; Heras et al.\ 2002) and its strength increases with decreasing temperature (i.e., later spectral types).  The 2--0, 3--1, and 2--4 bands of the $^{13}$C$^{16}$O isotope are also present in these giant stars  due to the larger CO absorption and decreased $^{12}$C/$^{13}$C ratio (Hinkle \& Wallace 1996).  Additionally, the Ca multiplet and the Fe line in the 2.261-2.267 \micron\ region strengthen somewhat in spectral types later than early G.  The Mg line near 2.28 \micron\ does not show much variation with temperature (Wallace \& Hinkle 2006), and is present at similar strengths in all our spectra.  In order to classify our background stars, we compared our observed spectra to the standard star spectra available in the IRTF Spectral Library which contains 57 G0 through M10 giant-star spectra \nocite{cushing2005,rayner2009} (Cushing, Rayner, \& Vacca 2005; Rayner et al.\ 2009).   The standard-star spectra were first  reddened by applying the  Indebetouw et al. (2005) extinction law  from 1.25 to 8.0 \micron, assuming constant extinction longward of 8 \micron.  We assume $A_K/A_V=0.09$, appropriate for R=3.05 \nocite{whittet2003} (Whittet 2003).  The visual extinction, $A_V$, was deduced based on $E(J-K)$, assuming average intrinsic colors for late G through early M giants from \nocite{tokunaga2000} Tokunaga et al. (2000) as a first best guess, and $A_V = 5.3 * E(J-K)$, appropriate for dense clouds \nocite{whittet2001} (Whittet et al. 2001).  Using the deduced extinction curve and taking advantage of the sensitivity of the CO lines to temperature, we determined the spectral type for each source (Fig.~\ref{fig:spectralclass}; Table~1).  The uncertainty is no more than $\pm1$ in subtype.  Following the classification, $E(J-K)$ was recalculated using the intrinsic colors for the appropriate spectral type; these values are given in Table~1.

%% FIGURE 2: color-color figure

%%% TABLE 2
\begin{deluxetable*}{cccccccccc}
\tabletypesize{\scriptsize}
%\rotate
\tablecaption{Peak Optical Depths of Various Ice and Dust Components\label{table:taus}}
\tablewidth{0pt}
\tablehead{
\colhead{Source ID} &
 \colhead{$\tau_{3.05}$} & 
 \colhead{$\tau_{3.47}$} &
 \colhead{$\tau_{6.0}$} & 
\colhead{comp. 1} & \colhead{comp. 2} & \colhead{comp. 3} & \colhead{comp. 4} & 
                                    \colhead{comp. 5} & 
\colhead{$\tau_{9.7}$}}
\startdata
Q21-1 & 1.45  & 0.065    & 0.19 & 0.04  & 0.04  & 0.07 & 0.06 & 0.00  & $0.602\pm0.002$ \\
Q21-2 & 0.57  & 0.030    & 0.09: & 0.10: & 0.04: & 0.07: & 0.06 &  0.00 & $0.534\pm0.110$ \\
Q21-3 & \nodata & \nodata & 0.04: & \nodata & \nodata & \nodata & \nodata & \nodata & $0.444\pm0.026$ \\
Q21-4 & 0.31 & $<0.02$ & 0.01: & \nodata & \nodata & \nodata & \nodata & \nodata & $0.281\pm0.082$ \\
Q21-5 & 0.30 & $<0.01$ & 0.04 & 0.02 & 0.02 & 0.02 & 0.02 &  0.00 & $0.259\pm0.006$ \\
Q21-6 & 1.17 & 0.055 & 0.17 & 0.05 & 0.04 & 0.08 & 0.07 & 0.00 & $0.689\pm0.008$ \\
Q22-1 & 0.82 & 0.048 & 0.11 & 0.03 & 0.04 & 0.04 & 0.04 & 0.00 & $0.509\pm0.006$ \\
Q22-3 & \nodata & \nodata & $<0.01$ &  \nodata & \nodata & \nodata & \nodata & \nodata & $<0.15$ \\
Q23-1 & 0.24 &  $<0.04$     & 0.06 & 0.03 & 0.03 & 0.03 & 0.03 & 0.00 & $0.297\pm0.017$ \\
Q23-2 & \nodata & \nodata & 0.05 & 0.03 & 0.03 & 0.05 & 0.05 & 0.00 & $0.379\pm0.026$ \\
\enddata
\end{deluxetable*}

\subsection{Determination of Continuum and Optical Depth Profiles}
\label{sec:continua}

The determination of an underlying stellar continuum is one of the biggest sources of uncertainty when deducing ice and dust absorption profiles.  Polynomial continua are somewhat arbitrary, but are often used when the true stellar continuum is not known and/or cannot be easily modeled.  Since we have photometry and spectroscopy over a wide wavelength region and we are able to determine the spectral types of our program stars with relative accuracy (\S\ref{sec:sptype}), we are able to fit reddened photospheric models from \nocite{decin2007} Decin \& Eriksson (2007).  These models cover the spectral region from 2 to 40 \micron\ and are available for  K1, K2 and K5 giants.   Thus, after the spectral type was determined based on the $K$-band spectrum, the Decin \& Eriksson model closest to that spectral type was chosen.  The model was reddened using the Indebetouw et al.\ (2005) extinction law as described above.   After convolving the reddened model flux with the appropriate photometry point, the model was normalized to the data.  In most cases, the reddened model was normalized to the $K_s$ band, except for   Q22-3 and Q21-4, where the IRAC4 and IRAC3 photometry, respectively, was used to normalize the reddened model.  The reddened models provide reasonable continua over the entire spectral energy distribution (Fig.~\ref{fig:continua}). The optical depth spectra were computed using  $\tau = -\ln F_{\nu}/F_{\rm continuum}$, and are presented in Fig.~\ref{fig:opticaldepths}.

\section{Ice and Dust Absorption Features} 
\label{sec:iceprofiles}
Several dust and ice absorption features are apparent in the optical depth spectra (Figure \ref{fig:opticaldepths}): the 3.0 \micron\ H$_2$O-ice and associated wing, the 6.0 \micron\ and 6.8 \micron\ ice/dust features, and the 9.7 \micron\ silicate feature.  Figure~\ref{fig:coice} shows the CO-ice absorption profile for  Q21-1.  

%% FIGURE 3: spectral classification
\begin{figure}
\figurenum{3a}
\includegraphics[angle=0,scale=0.45]{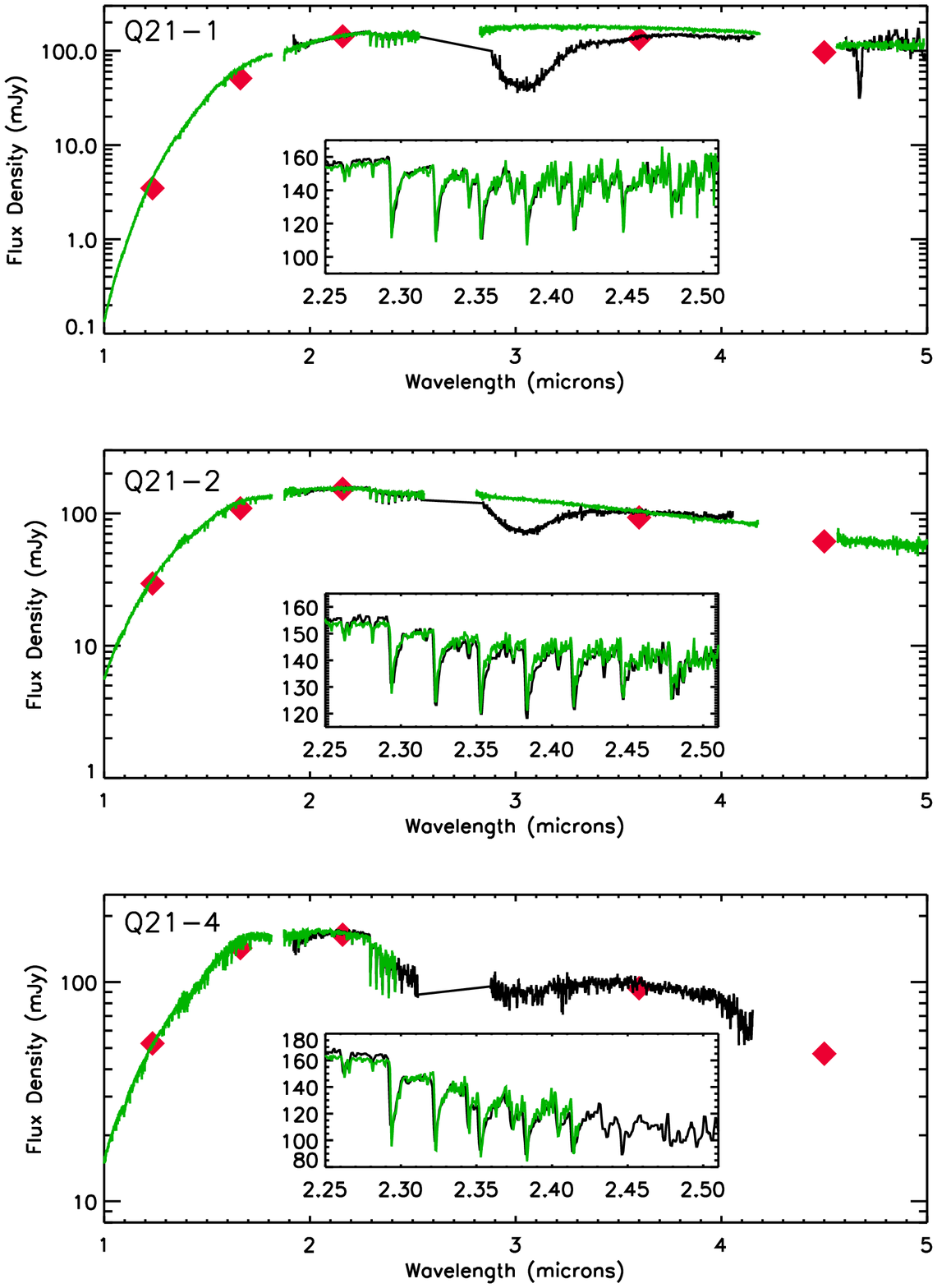}
\caption{Spectral energy distributions for the IC 5146 field stars with available
IRTF-SpeX data. 2MASS photometry (diamonds), SpeX spectra of IC 5146 field star (black line), SpeX standard star (green line)}
\label{fig:spectralclass}
\end{figure}

\begin{figure}
\figurenum{3a}
\includegraphics[angle=0,scale=0.45]{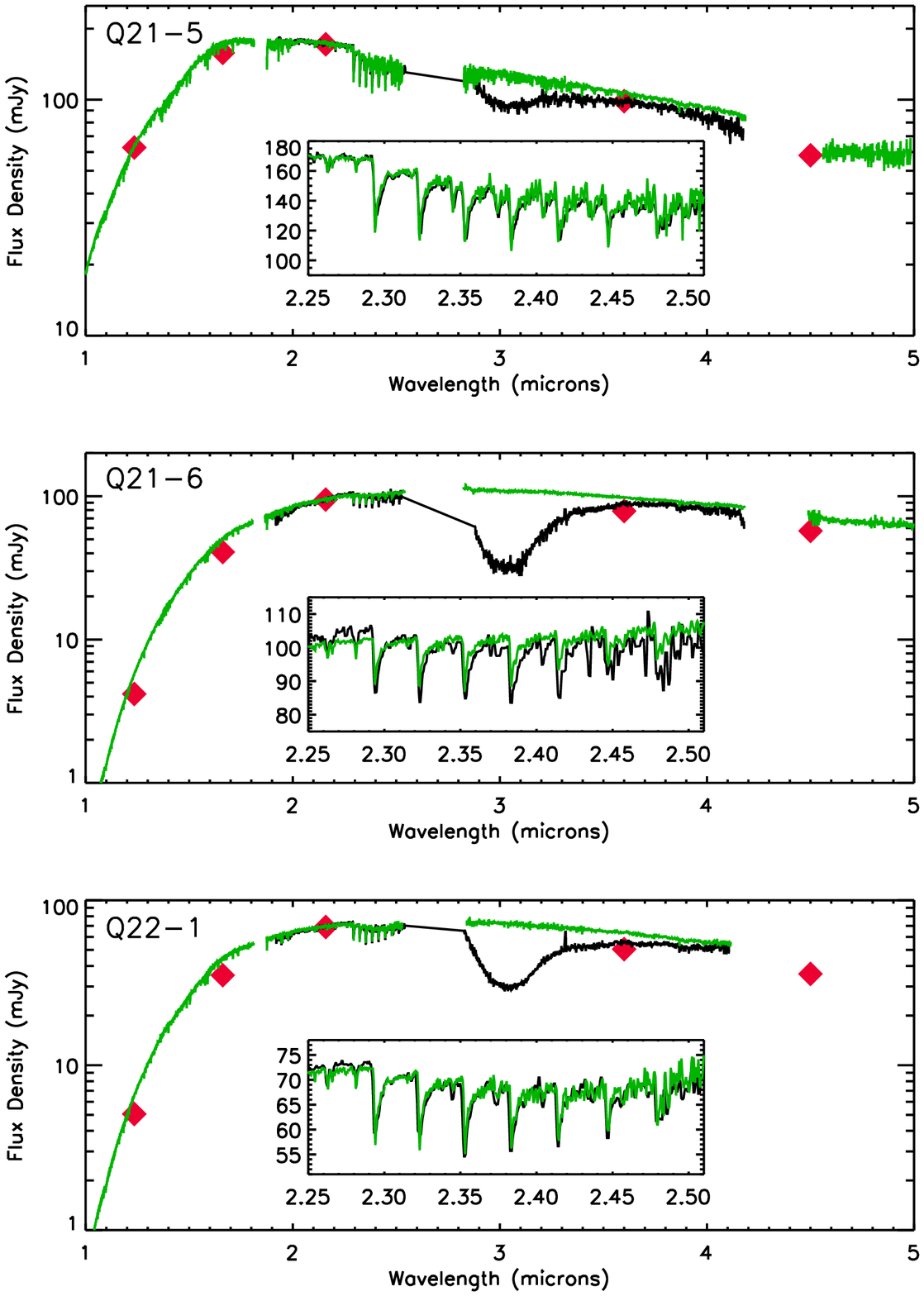}
\caption{continued}
\end{figure}

\begin{figure}
\figurenum{3a}
\includegraphics[angle=0,scale=0.45,trim = 0mm 160mm 0mm 0mm,clip]{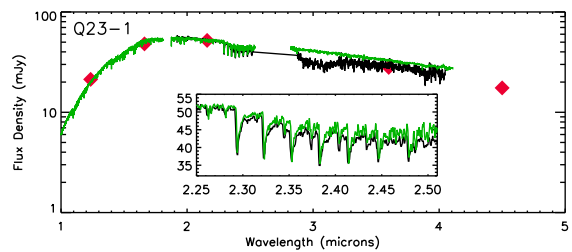}
\caption{continued}
\end{figure}

\begin{figure}
\figurenum{3b}
\includegraphics[angle=0,scale=0.45]{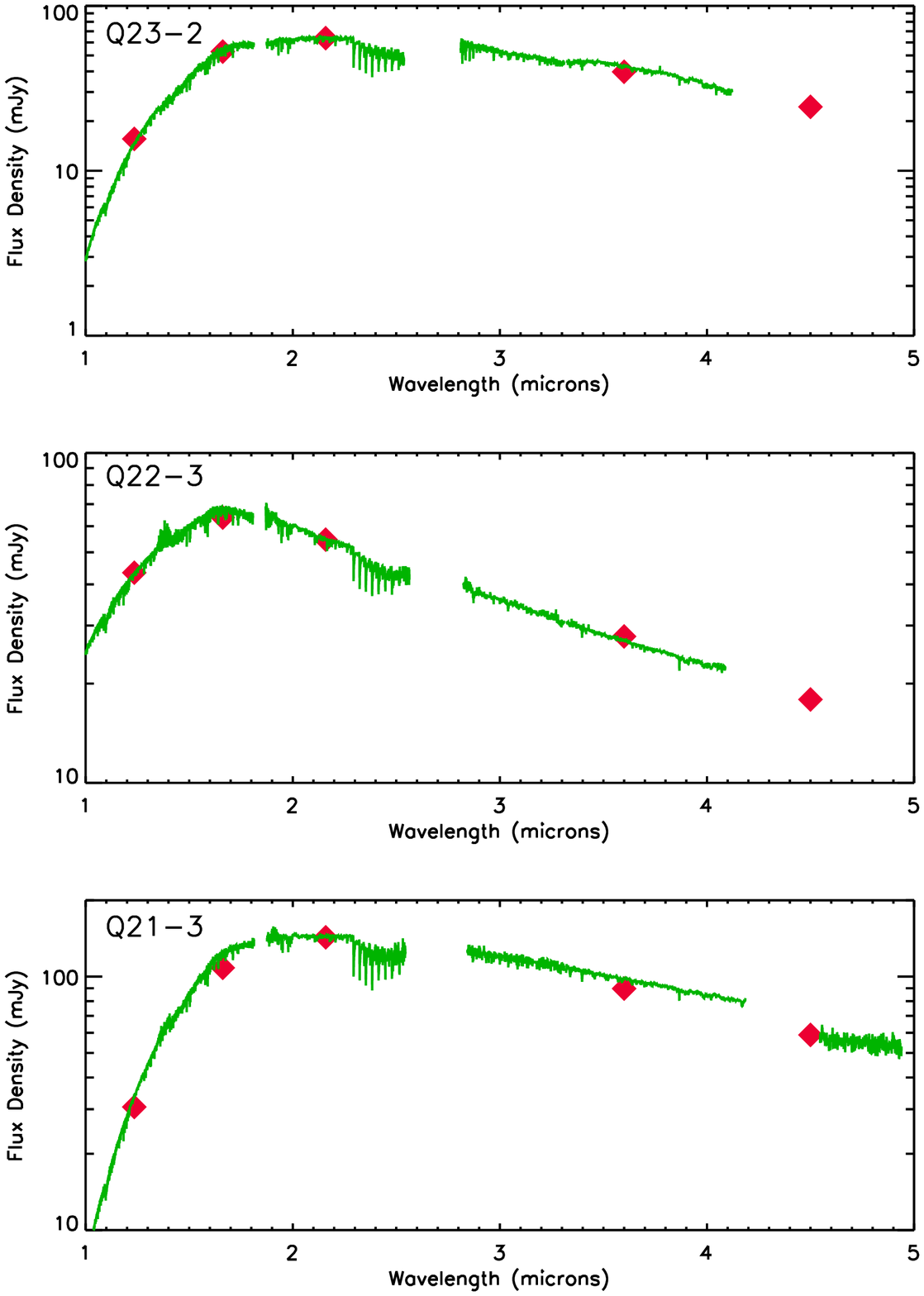}
\caption{IC 5146 field stars with no available SpeX data.}
\label{fig:spectralclass_nospex}
\end{figure}

\subsection{\water-ice and Long-wavelength Wing}
 \label{sec:longwavewing}
Water-ice is the primary ice-mantle constituent in dense clouds and its presence is evident through detection of  its fundamental O-H stretching-mode absorption feature at 3.0 $\mu$m and bending and libration modes at 6.0 and 13 \micron, respectively.  In laboratory ice analogs, the peak depth of the 6.0 \micron\ ice absorption feature is 7 to 9 times weaker than the  3.0 \micron\ absorption, and in the astronomical spectra absorption at 6.0 \micron\ is a blend of \water-ice with other ice components (see below).  The libration band is also weak and blended on the shoulder of the broad 9.7 \micron\ silicate absorption feature.  The presence of alcohols and other compounds containing OH groups in the ice can contribute to the 3.0 \micron\ absorption band and should be taken into account for those lines of sight that have \methanol\ abundances greater than 20\% relative to \water.  Since the methanol abundance for the objects studied here is on the order of a few percent, the contribution of alcohols to the 3.0 \micron\ absorption can be neglected.  Since the 3.0 \micron\ band doesn't usually suffer from severe issues of blending and is the strongest of the \water-ice absorption features, we use it to deduce the \water-ice column density as well as the ice ``threshold'' extinction.

\begin{figure}
\figurenum{4}
\includegraphics[angle=0,scale=0.45]{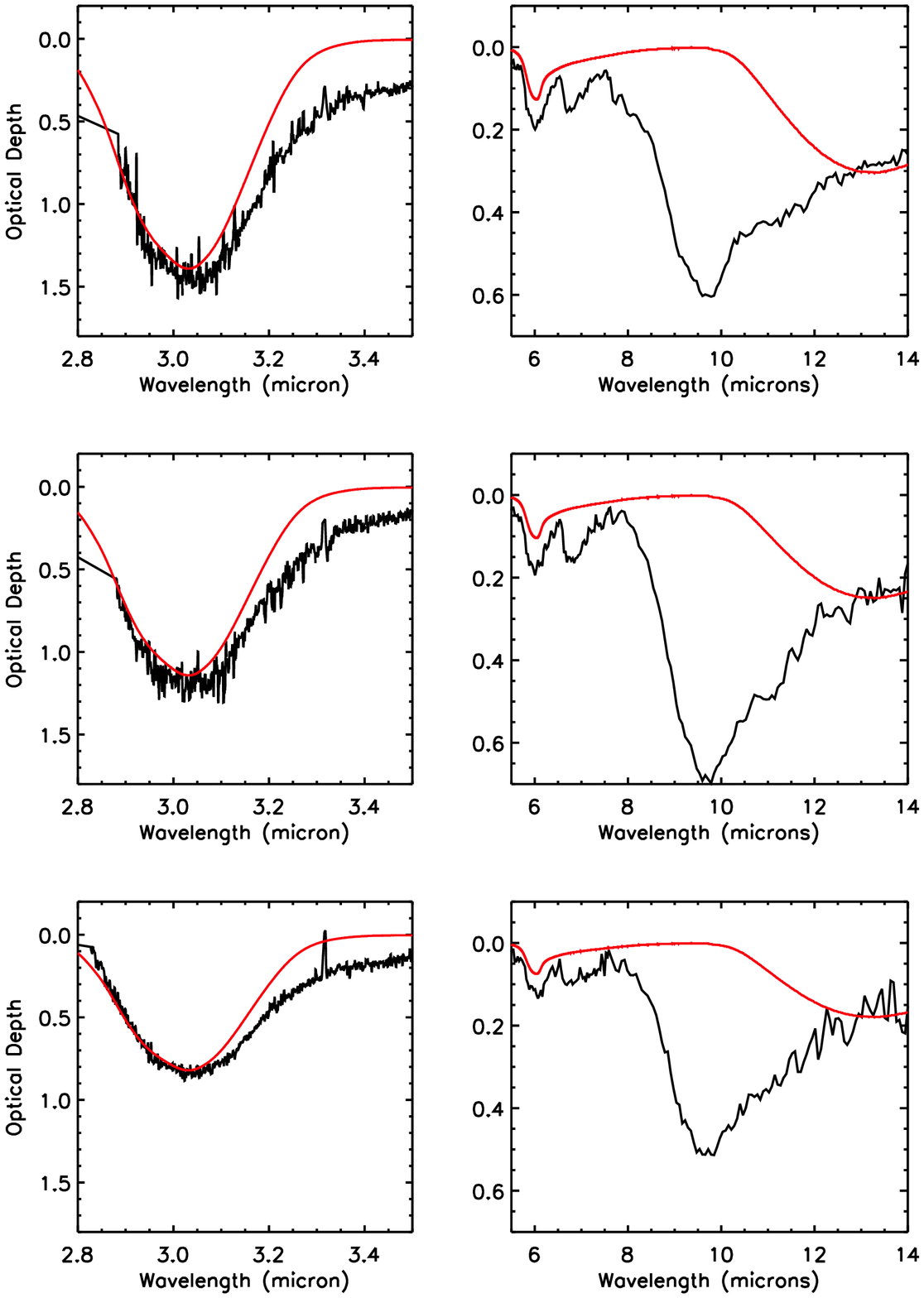}
\caption{Optical depth profiles for the ice and dust features based on the reddened photosphere model matches as shown in the previous figure.  Laboratory data (red lines) for H$_2$O ice at 10K are overlaid on the observational data (black lines). Sources are (from top to bottom) Q21-1, Q21-6, Q22-1.}
\label{fig:opticaldepths}
\end{figure}

Fig.~\ref{fig:opticaldepths} shows the laboratory spectrum of 10\,K H$_2$O-ice \nocite{hudgins1993} (Hudgins et al.\ 1993) overlaid on the optical depth spectra.  Compared to the laboratory ice analog, the observed 3 \micron\ ice profile has ``excess'' absorption longward of 3.05 \micron\ and, in some cases, weak excess absorption at about 2.95 \micron.  Smith et al. (1993) \nocite{smith1993} noticed the latter feature in Taurus field stars on later spectral types, and therefore attributed it to a photospheric or circumstellar feature.  However, for the IC~5146 stars, the spectra with the most apparent excess at 2.96 \micron\ are late G and early K giants, so a stellar origin is less likely.   Also present in these spectra is ``excess'' absorption in the 3.2 to 3.6 \micron\ region, known as the (enigmatic) ``long-wavelength wing'' \nocite{smith1993,baratta1990} (Baratta \& Strazzulla 1990; Smith et al.\ 1993).  Ammonia (\ammonia) ice related species such as ammonium-hydrates may account for some of this absorption  as well as the excess at 2.96 \micron\ \nocite{chiar2000,gibb2001} (e.g. Chiar et al.\ 2000; Gibb et al.\ 2001).  Some of the wing absorption may be accounted for by scattering in a model where spherical grains are coated with thick mantles or only the largest grains have mantles (Smith et al.\ 1993).   Discrete features in the 3.2 to 3.6 \micron\ region contribute to the absorption: a feature at 3.47 \micron\ is  attributed to hydrocarbon-containing ices  \nocite{chiar1996,brooke1996, brooke1999} (Chiar et al. 1996; Brooke et al. 1996, 1999) and \methanol-ice exhibits a narrow absorption feature at 3.54 \micron\ \nocite{allamandola1992} (Allamandola et al.\ 1992).  However, in quiescent dense clouds the abundance of \methanol\ relative to \water-ice is less than 5\%, so it does not contribute significantly (Chiar et al.\ 1996).  Boogert et al.\ (2011) find the \methanol\ abundance to be $\sim7$\% for the isolated dense cores probed by background stars.  Thus, much of the ``wing'' absorption remains a mystery.  

%% FIGURE 4: optical depth profiles
%% 3 pages (leave out spectrum with silicate=limit

\begin{figure}
\figurenum{4}
\includegraphics[angle=0,scale=0.45]{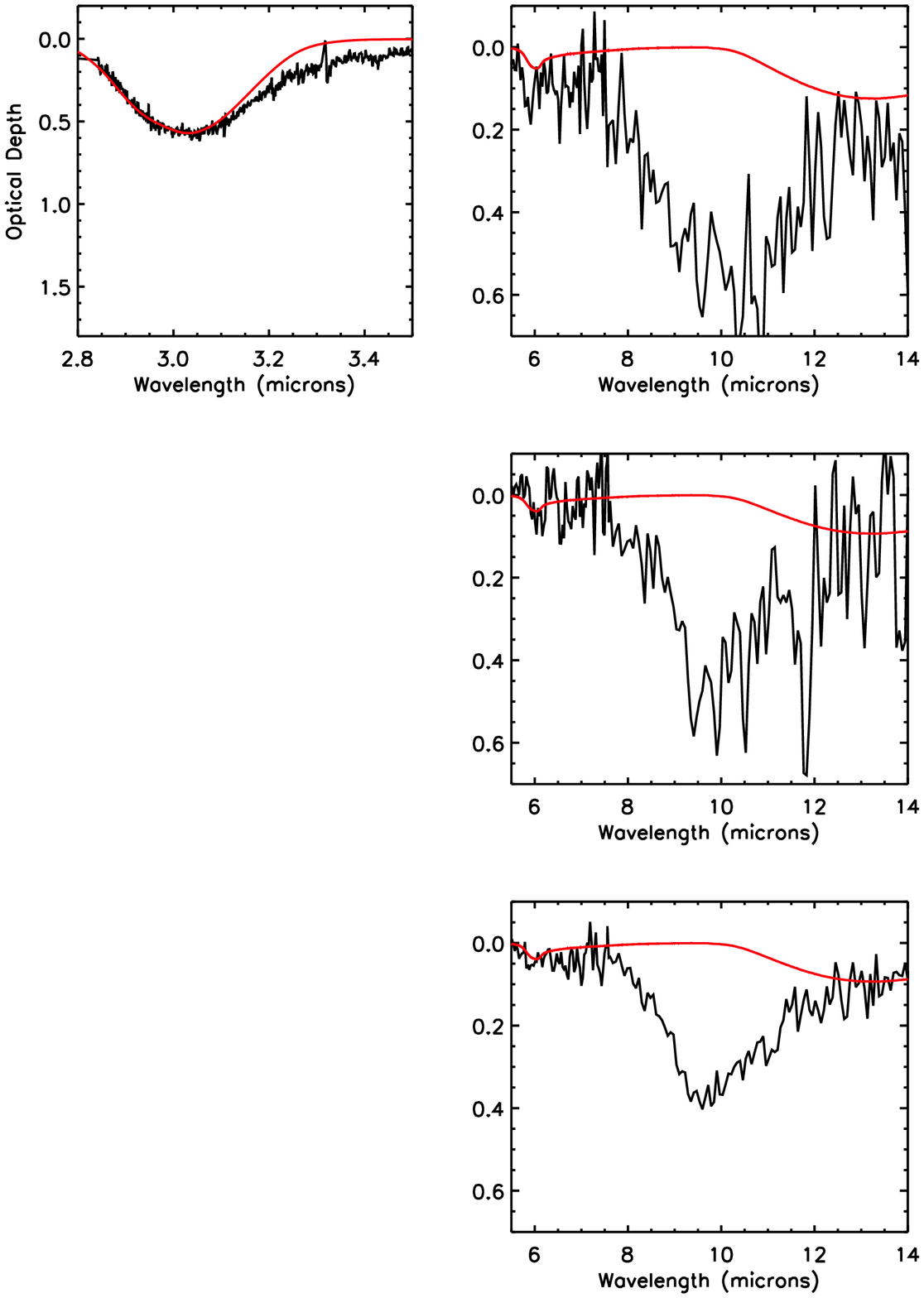}
\caption{Profiles for (from top to bottom) Q21-2, Q21-3, Q23-2.}
\end{figure}

\begin{figure}
\figurenum{4}
\includegraphics[angle=0,scale=0.45]{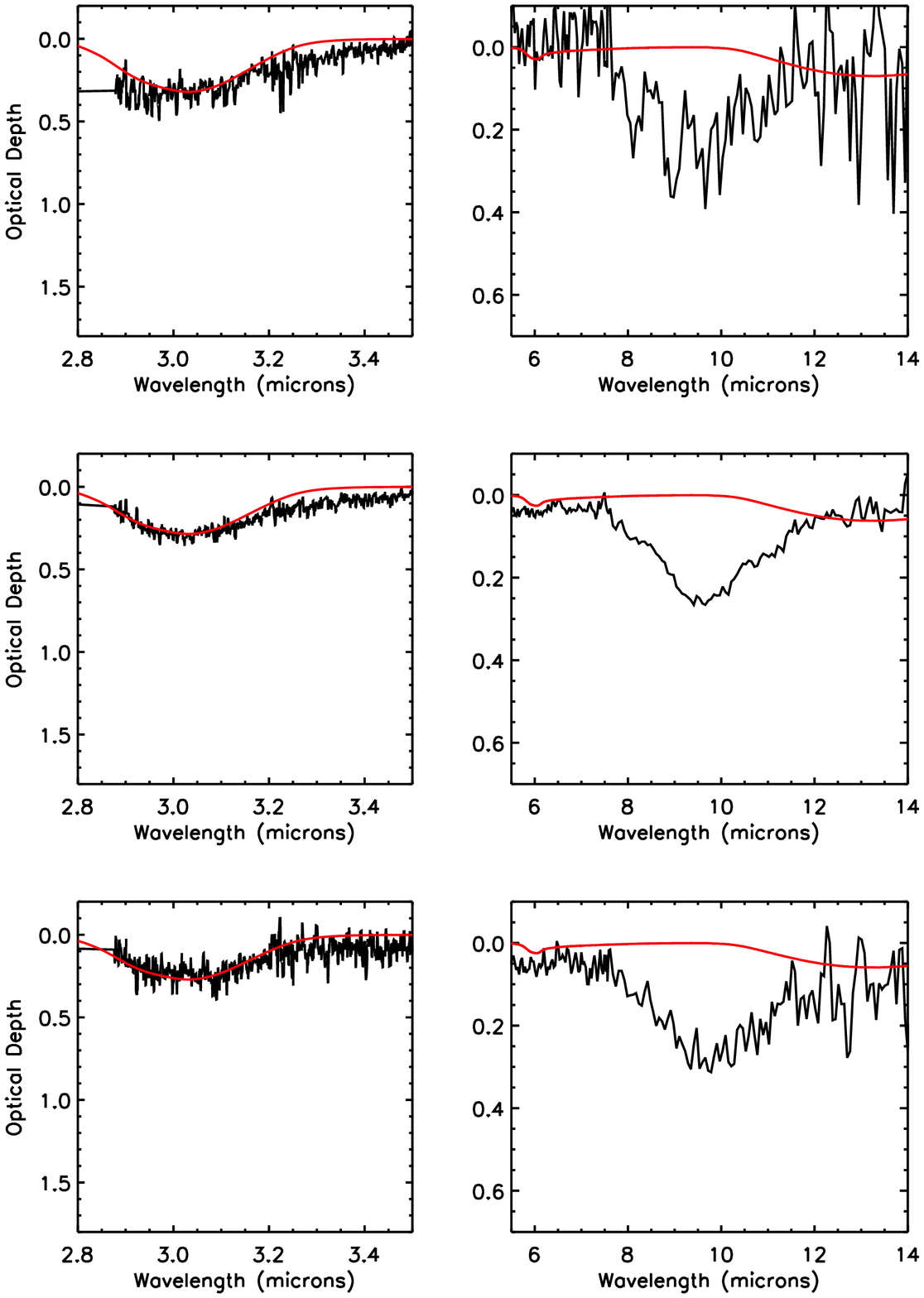}
\caption{Profiles for (from top to bottom) Q21-4, Q21-5, Q23-1.}
\end{figure}

\begin{figure}
\figurenum{5}
\includegraphics[angle=0,scale=0.35]{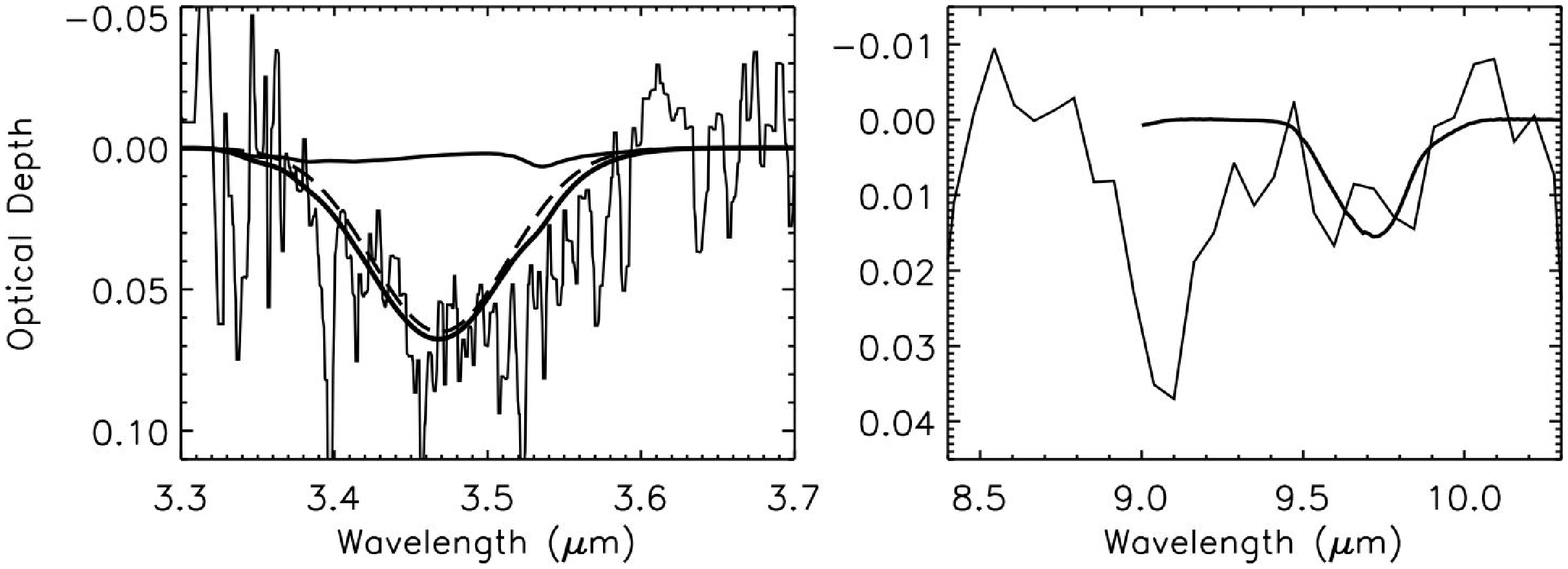}\vspace*{0.1in}
\includegraphics[angle=0,scale=0.35]{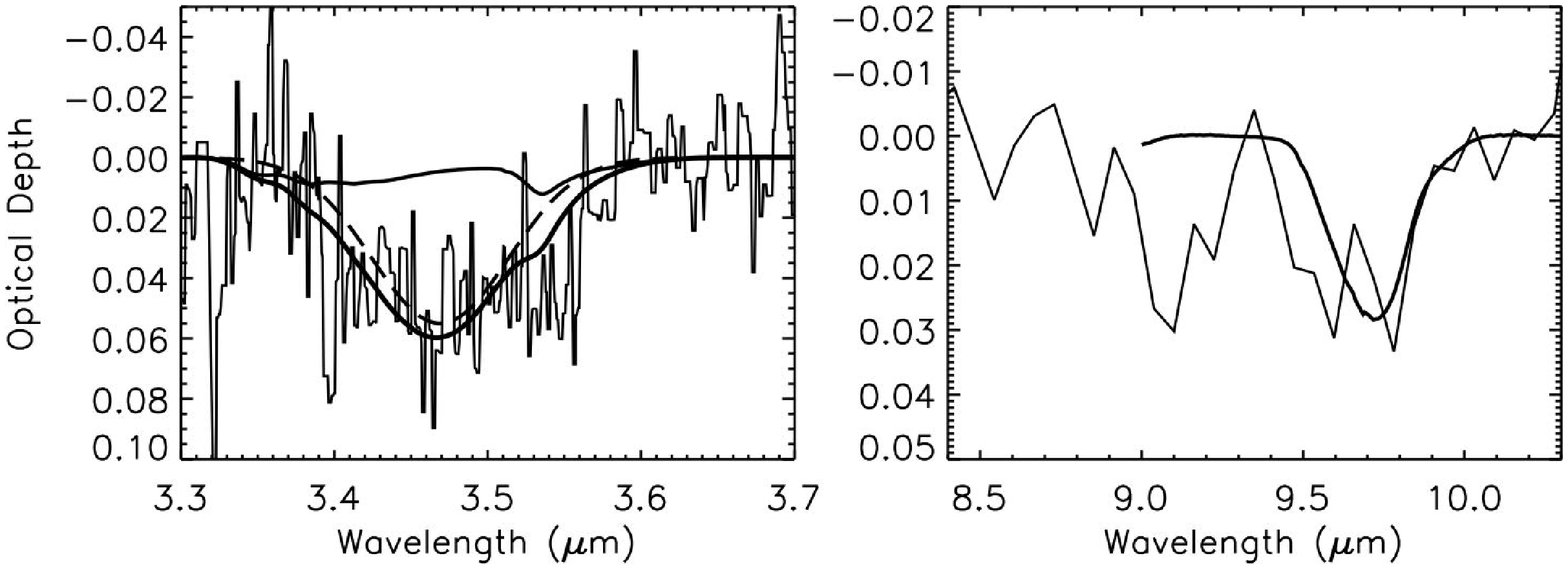}
\caption{Absorption in the long-wavelength \water-ice wing [left panel] and 8.5 to 10 \micron\ region after fitting local continua to remove \water-ice and silicate absorption (see \S\ref{sec:longwavewing} and \ref{sec:traceices} for details) for Q21-1 [top] and Q21-6 [bottom]. The left panel shows a Gaussian representing the average 3.47 \micron\ feature profile as determined by Brooke et al.\ 1999 (dashed line), the weak \methanol\ ice feature (solid line) from laboratory data (Gerakines et al.\ 1995, 1996), and coaddition of those two features (deepest smooth line).  The contribution of \methanol\ at 3.54 \micron\ is constrained by the strength of the maximum allowed \methanol\ feature detected at 9.75 \micron\ (left panel; laboratory data shown as smooth curve).}
\label{fig:icewing}
\end{figure}

Four of our sources have a detectable 3.47 \micron\ feature.  The optical depth, $\tau_{3.47}$ was calculated using a local third degree polynomial to determine the continuum across the 3.2 to 3.8 \micron\ region, avoiding the 3.30 to 3.33 \micron\ region (where the telluric cancellation is poor) and the 3.4 to 3.6 \micron\ region where the ice absorption is expected to be present. We determined $\tau_{3.47}$ using the average feature width ($\Delta\lambda=0.105\pm0.004$\micron) and central wavelength ($\lambda_0=3.469\pm0.002$ \micron) determined by \nocite{brooke1999} Brooke et al.\ (1999).   The results are given in Table~2.  In addition, for Q21-1 and Q21-6, absorption in excess of the typical 3.47 \micron\ feature is present.   Fig.~\ref{fig:icewing} shows that this absorption is consistent with the \methanol\ amounts allowed by the observed 9.75 \micron\ feature in the spectra of these same sources.

%% FIG  5: wing and methanol at 9 mic region

In spite of the excess wing absorption,  the 3.0 \micron\ absorption feature itself is well-suited for calculating the \water-ice column density.  We estimate the \water-ice column by scaling a Gaussian function with $\lambda_0=3.04$ \micron\ with FWHM = 0.22 \micron.  On average, the column densities determined in this manner are about 10\% larger than those determined using the 10\,K \water-ice laboratory spectrum.  The resulting column densities are listed in Table~\ref{table:cds}.  The laboratory ice analog spectrum is also useful for assessing the contribution of pure \water-ice to the 6.0 and 13.0 \micron\ spectral regions.  In order to limit the contribution of ice to those spectral regions, we fitted the laboratory spectrum to the trough of the 3.0 \micron\ absorption and calculated a residual across the entire Spitzer IRS spectrum.  This calculated residual was then used to analyze the 9.7 \micron\ silicate absorption profiles (\S\ref{sec:silicates}) and the absorption features in the 5 to 8 \micron\ region (\S\ref{sec:weakfeatures}).    

%% FIGURE 6: Correlation h2o ice vs Av
\begin{figure}
\figurenum{6}
\includegraphics[angle=90,scale=0.35]{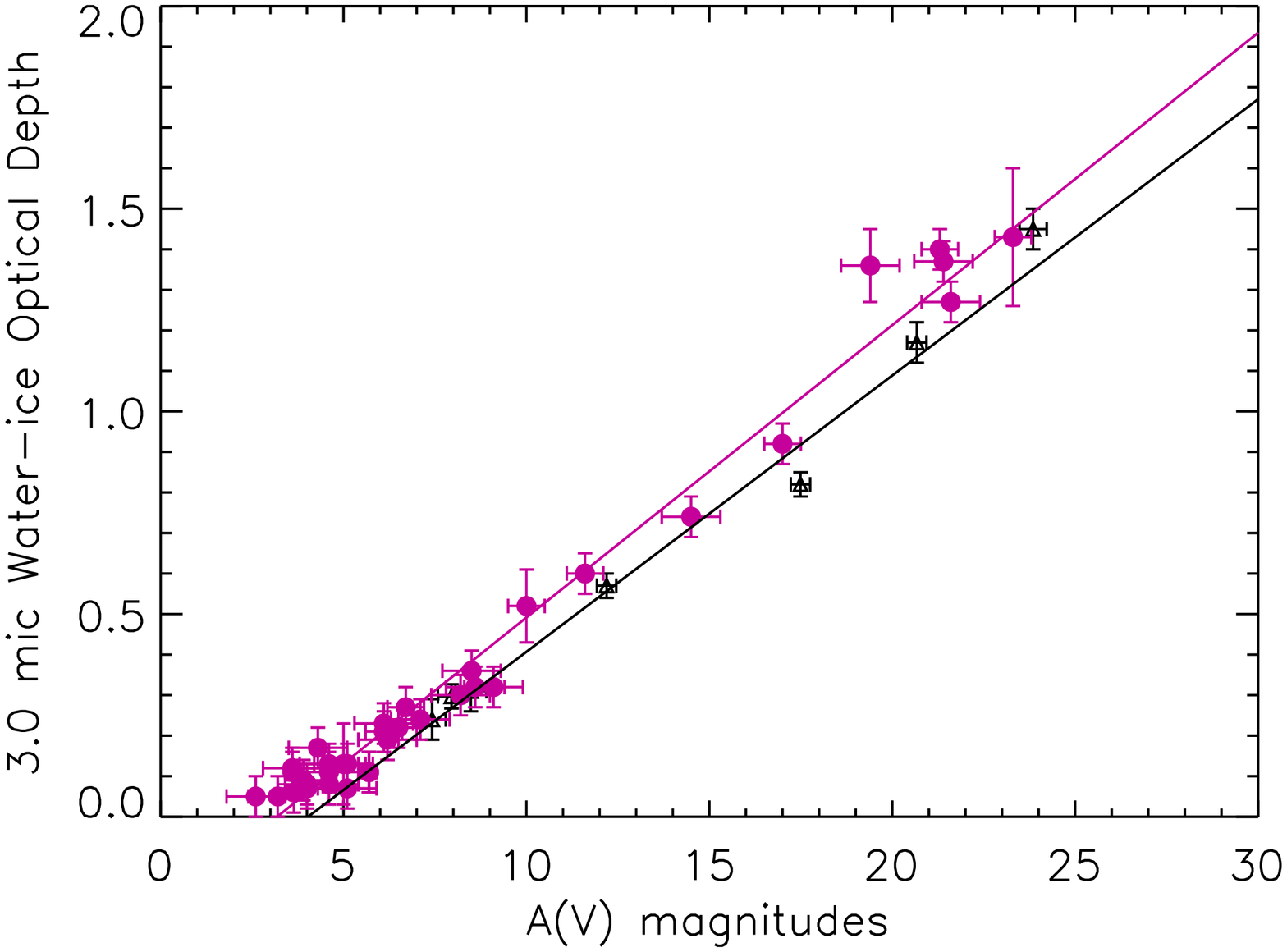}
\caption{Optical depth of the 3.0 \micron\ H$_2$O-ice feature versus visual extinction for field stars behind the IC~5146 (this work, black symbols) and Taurus (Whittet et al.\ 2001, purple symbols) dark clouds.    \av\ values for the IC 5146 background stars are based on \ejk\ as determined from the reddened photospheric model fit and an appropriate extinction law (see \S\ref{sec:sptype} and \ref{sec:continua}).  Determination of \av\ for Taurus sources is  discussed in detail in Whittet et al.\ (2001).  A weighted least-squares fit to the IC 5146 data (black line) and Taurus (purple line) taking uncertainties in $\tau_{3.0}$ and $A_V$ into account results in threshold extinctions of $A_V=  4.02$ mag and $A_V=  3.19$ mag, respectively.  Accounting for the foreground extinction toward IC 5146 ($\sim0.8$ mag) brings the threshold \av\ for IC 5146 and Taurus into agreement.} 
\label{fig:icecorr}
\end{figure}

\subsubsection{\water-ice Threshold}
 \label{sec:threshold}
The ice threshold is the  \av\ value below which ice absorption is not detected and therefore icy mantles are not present  (Whittet 2003).  The threshold $A_V$ is generally higher for dense clouds that have active star formation and are forming intermediate- to high-mass stars, compared to quiescent clouds that form low-mass stars at a slow rate.   Volatile ice species, like CO-ice, whose presence is more sensitive to local temperature and density conditions tend to have higher threshold extinction \nocite{whittet1989,chiar1995} (Whittet et al.\ 1989; Chiar et al.\ 1995).  The Taurus cloud, generally used as a model for the measurement of ice thresholds in pristine environments since it has the greatest number of ice absorption measurements toward background stars, exhibits a threshold of $A_V= 3.2\pm0.1$ mag \nocite{whittet2001} (Whittet et al. 2001 and references therein).  For the more active Serpens cloud, the \water-ice threshold is higher at $A_V\sim6$ \nocite{eiroa1989} (Eiroa \& Hodapp 1989), and for the $\rho$ Ophiuchus cloud that is forming intermediate- to high-mass stars, the threshold is even higher at $A_V\sim13$ mag \nocite{tanaka1990} (Tanaka et al.\ 1990).  

For the IC 5146 cloud, we calculate the \water-ice threshold extinction by means of a linear least-squares fit including the uncertainties (error bars) in both \av\ and $\tau_{3.0}$ , resulting in a threshold of $A_V=4.03\pm0.05$ mag  (Fig.~\ref{fig:icecorr}).  For consistency, we repeat the fit to the Whittet et al.\ (2001) data while taking the uncertainties in both $A_V$ and $\tau_{3.0}$ into account, resulting in threshold extinction, $A_V=3.19\pm0.07$.  As demonstrated by the correlation lines in Fig.~\ref{fig:icecorr}, the slopes of the IC 5146 and Taurus correlation lines are similar: $0.072\pm0.002$  and $0.068\pm0.003$, respectively, suggesting that \water-ice mantle growth with increasing extinction is similar in both clouds.  This is in line with calculations that show that once the critical threshold \av\ is reached -- i.e., the first mono-layer of ice is formed on the silicate or carbonaceous substrate -- the local infrared radiation field has little effect on subsequent ice layer growth due to strong H-bonding between neighboring \water-ice molecules \nocite{williams1992} (Williams et al.\ 1992).   The apparent increase of the threshold $A_V$ in IC 5146 is likely to be an effect of foreground extinction given the distance of 950 pc to IC 5146 (\S1).  In fact, Neckel \& Klare (1980) \nocite{neckel1980} estimate the foreground extinction in the direction of IC 5146 to be $\sim0.8$ mag, comparable to the difference in the threshold extinction between IC ~5146 and Taurus. 

%IC 5146 is also associated with an \ion{H}{2} region, though the region probed by the lines of sight discussed here are some 20 pc away from that region.  (It seems unlikely to me that the HII region is close enough to ``process'' the ice in our lines of sight.  What do you think?)

%% FIG 7: Q21-1 CO ice profile
\begin{figure}
\figurenum{7}
\center\includegraphics[angle=90,scale=0.35]{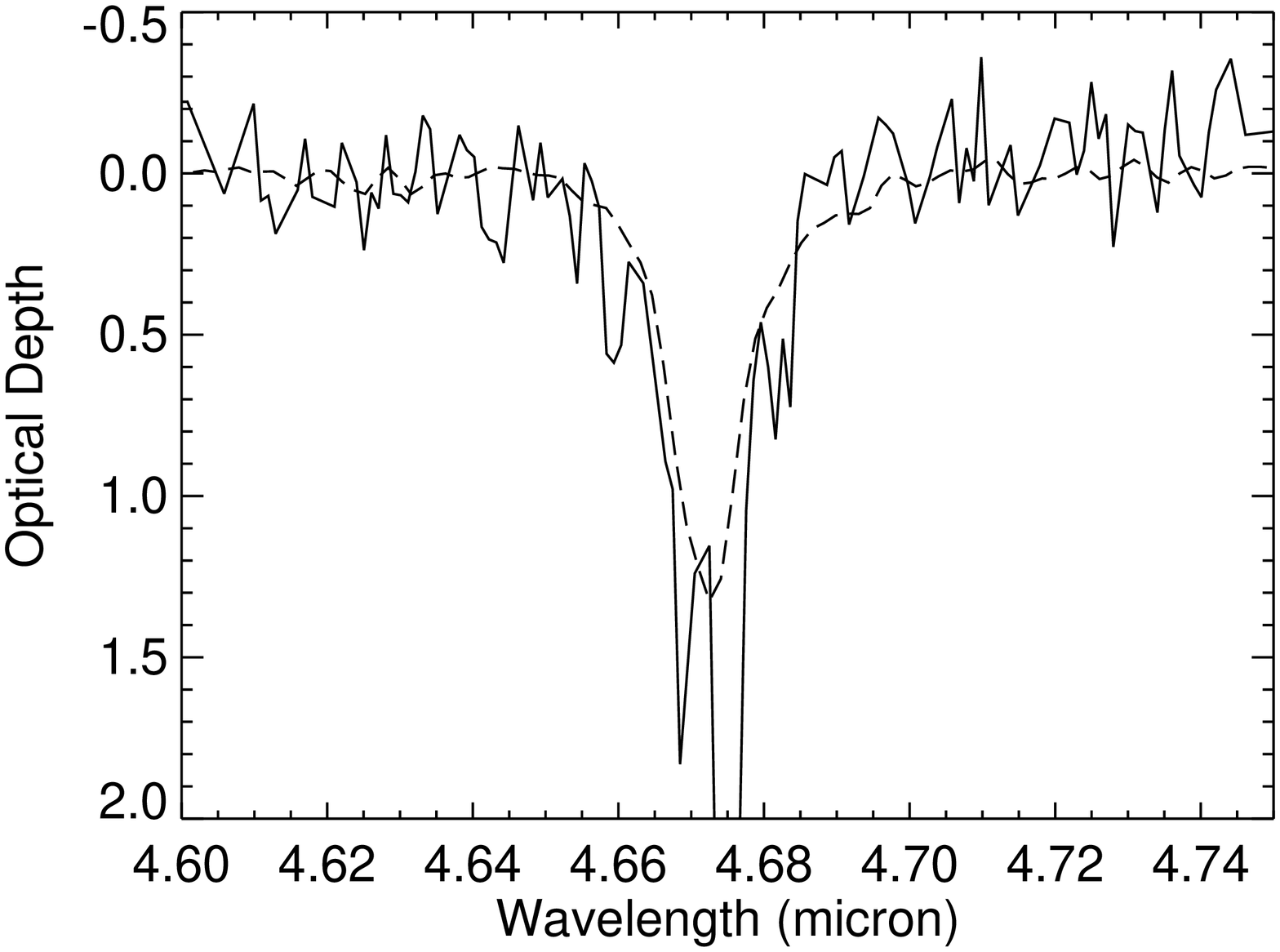}
\caption{Optical depth spectrum in the region of the CO ice absorption feature for Q 21-1 (solid line).  The (scaled) profile for the Taurus field star, Elias 16 is also shown for comparison (dashed line).  See text for details.}
\label{fig:coice}
\end{figure}

\subsection{CO and CO$_2$ Ice}
Toward field stars behind dense clouds, the more volatile CO-ice ranges in abundance  between 10 and 40\% relative to \water-ice  \nocite{chiar1995,whittet2009} (Chiar et al. 1995; Whittet et al.\ 2009).   We were able to acquire an M-band spectrum with IRTF-SpeX for one of the IC 5146 sources, Q21-1.  Figure~\ref{fig:coice} shows the 4.67 \micron\ CO ice absorption profile with the spectrum of the Taurus field star Elias 16 \nocite{chiar1995} (Chiar et al.\ 1995) overlaid for comparison.  Many of the sharp absorption peaks in the Q21-1 spectrum that dip below the Elias 16 spectrum are due to (unresolved) gas phase lines.  Since the resolution is not high enough to model these lines, we use the relatively smooth Elias 16 spectrum to estimate the CO ice column density.   The Elias 16 CO profile, and therefore also its column density, is scaled by a factor of 1.05 to give a CO ice column density  for Q21-1 equal to $6.83\times10^{17}$ cm$^{-2}$ (Table~\ref{table:cds}).  This results in CO/\water\ column density ratio equal to 0.27, in the same range found for other dense cloud lines of sight.

The bending mode of CO$_2$ occurs at 15.3 \micron\ and is contained in the IRS short-high spectrum for Q21-1 previously presented by \nocite{whittet2009} Whittet et al. (2009).  They found that the profile shape of the 15.3 \micron\ CO$_2$ feature is unvarying in the dark clouds sampled (Taurus, IC 5146, Serpens) despite some scatter in the abundance relative to \water-ice which may be due differences in how the elemental oxygen is distributed in the main oxygen-containing ice species (\water, CO, CO$_2$).  They find that the CO$_2$ abundance relative to \water\ is $\sim0.2$ for Taurus and $\sim0.3$ for IC 5146 (Q21-1).

%% FIG 8: silicate profile comparison
\begin{figure}
\figurenum{8}
\center\includegraphics[angle=90,scale=0.35]{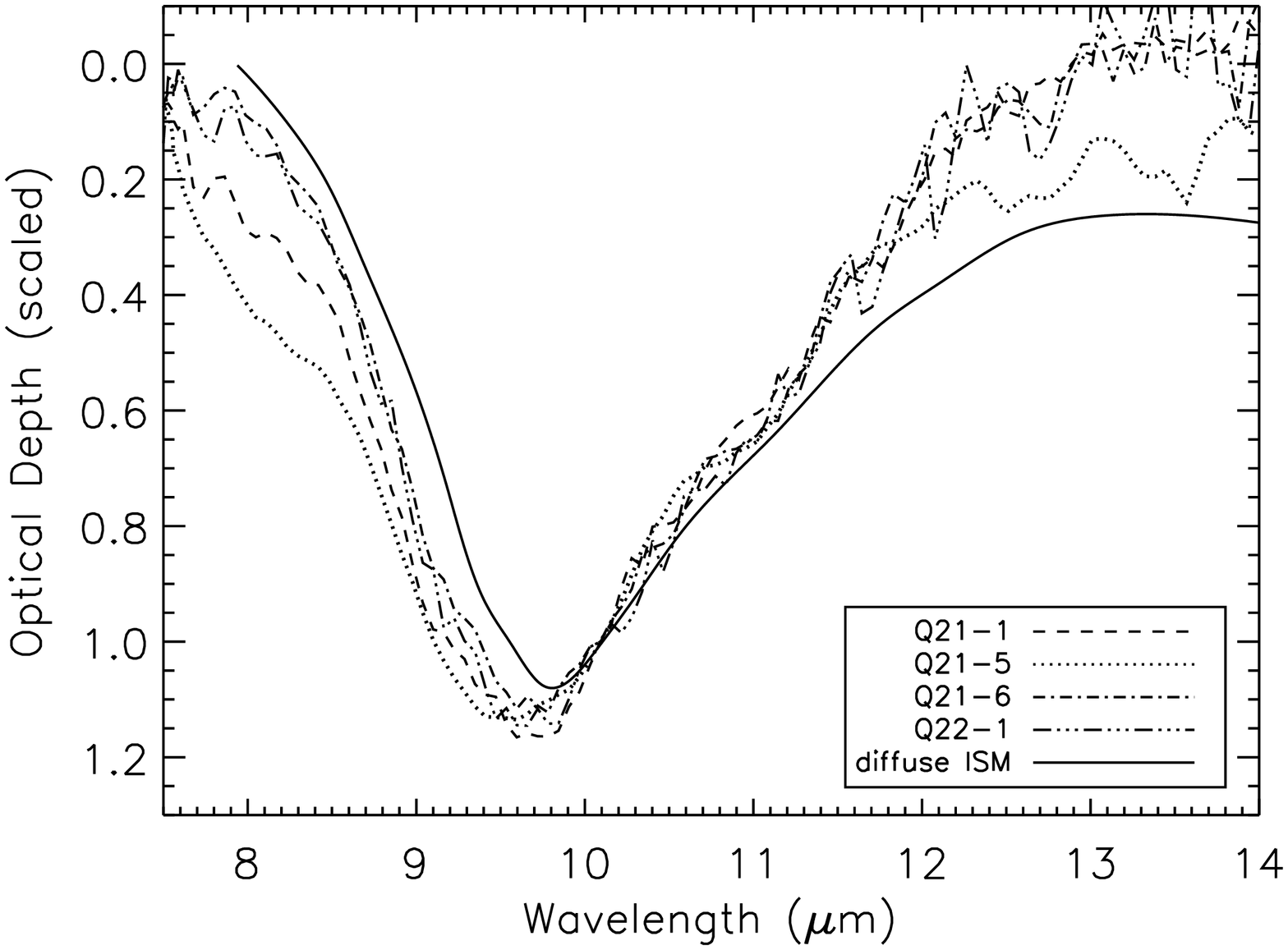}
\caption{Comparison of the silicate profiles for highest extinction sources Q21-1, Q-21-6, Q22-1 (dashed lines), the low-extinction source Q21-5 (dotted line) and the diffuse ISM as probed by WR 98a (solid line; Chiar et al.\ 2006).  The spectra are normalized at 10.1 \micron. Small differences are noticeable on the blue wing of the feature; these may be due to photospheric  absorption in the field stars.  For the high extinction sources, moderate excess absorption at the bottom of the feature ($\sim9.8$ \micron) is due to \methanol-ice absorption (see Fig.~\ref{fig:silice}).   }
\label{fig:silprofile}
\end{figure}

\subsection{Silicates}
 \label{sec:silicates}
Differences in the profile shape of the interstellar silicate feature observed in dense clouds and the diffuse ISM have been noted previously.  Generally, emissivity curves based on observations of the M supergiant $\mu$ Cephei \nocite{russell1975} (Russell et al.\ 1975) and Orion's Trapezium region \nocite{forrest1975} (Forrest et al.\ 1975) are used to fit the observed absorption features in the diffuse ISM and dense clouds, respectively \nocite{roche1984,whittet1988,bowey1998} (e.g. Roche \& Aitken 1984; Whittet et al.\ 1988; Bowey et al.\ 1998).    The Trapezium silicate profile peaks at 9.6 \micron\ compared to 9.9 \micron\ for the $\mu$ Cephei profile.  Furthermore, \nocite{vanbreemen2010} van Breemen et al.\ (2010) find that the ``dense cloud'' silicate profiles differ substantially from the diffuse ISM profiles peaking at shorter wavelengths relative to the diffuse ISM profiles.  To illustrate the robustness of these differences,    we compare the IC 5146 field star profiles with the highest S/N to that of the diffuse ISM toward the Wolf-Rayet star, WR 98a \nocite{chiar2006} (from Chiar \& Tielens 2006) in Fig.~\ref{fig:silprofile}.  All profiles have been normalized to unity at 10.1 \micron.  The normalization wavelength was chosen to avoid regions that could be contaminated by photospheric or ice absorption.  This comparison shows that there are small differences in the dense cloud spectra on the blue side; these differences may be due to residual photospheric absorption (like SiO) that is not fully accounted for in the models.   In addition, there is ``excess'' absorption in the trough of the feature in the highly reddened sources due to \methanol- and \ammonia-ice absorption (see \S\ref{sec:traceices}).  Otherwise, the silicate profiles show little difference in the less- and more-reddened lines of sight in this small sample.  All the IC 5146 field star profiles are shifted blueward compared to the diffuse ISM profile.

%% FIG 9: correlation between tau 3 and tau 6 
\begin{figure}
\figurenum{9}
\includegraphics[angle=90,scale=0.35]{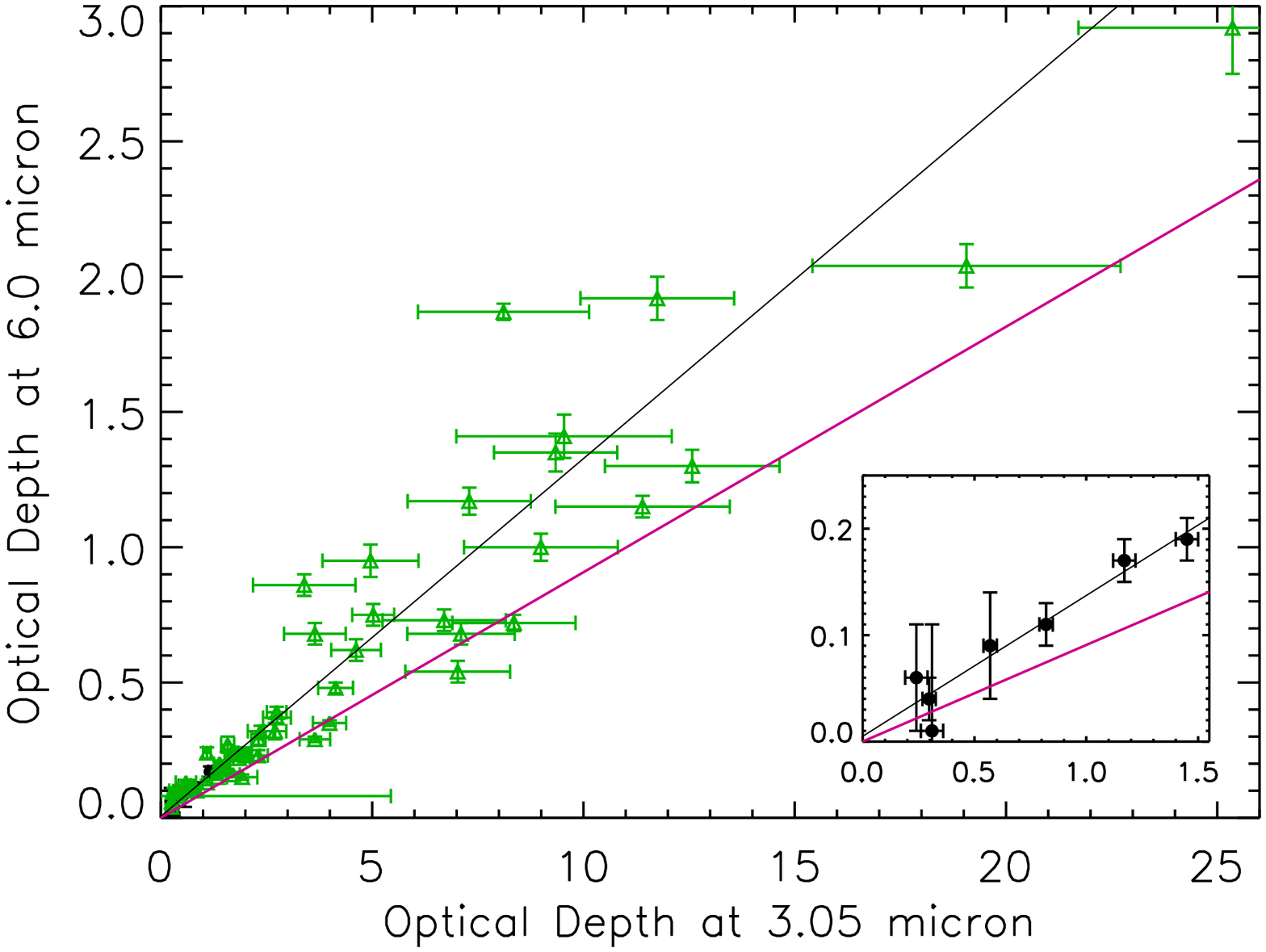}
\caption{The relationship between the optical depths of the 3.05 and 6.0 \micron\ features.   Data points are shown for the observed features in IC 5146 (black points) and low-mass YSOs (green points, Boogert et al.\ 2008).   The IC 5146 points are clustered in the lower left corner of the plot; the inset shows this region in more detail.    The purple line is the ratio, $\tau_{6.0}/\tau_{3.05}$, for laboratory \water-ice at 10K (Hudgins et al.\ 1993).  The black line is the least-squares fit to just the IC 5146 sources.}
\label{fig:corr36}
\end{figure}

\subsection{Absorption Features in the 5 to 8 \micron\ Region}
 \label{sec:weakfeatures}
Absorption features centered around 6.0 and 6.8 \micron\ are detected in six out of ten lines of sight in IC~5146 (see Table~2).   These features have been previously detected in the spectra of high-mass \nocite{schutte1996b,keane2001}(Schutte et al.\ 1996; Keane et al.\ 2001) and low-mass young stellar objects (YSOs; Boogert et al. 2008), dense clouds as probed by a few field stars (Knez et al. 2005) and dense cores as probed by field stars (Boogert et al.\ 2011).    

%% FIG 10: 5-8 mic components
\begin{figure}
\figurenum{10}
\includegraphics[angle=0,scale=0.45]{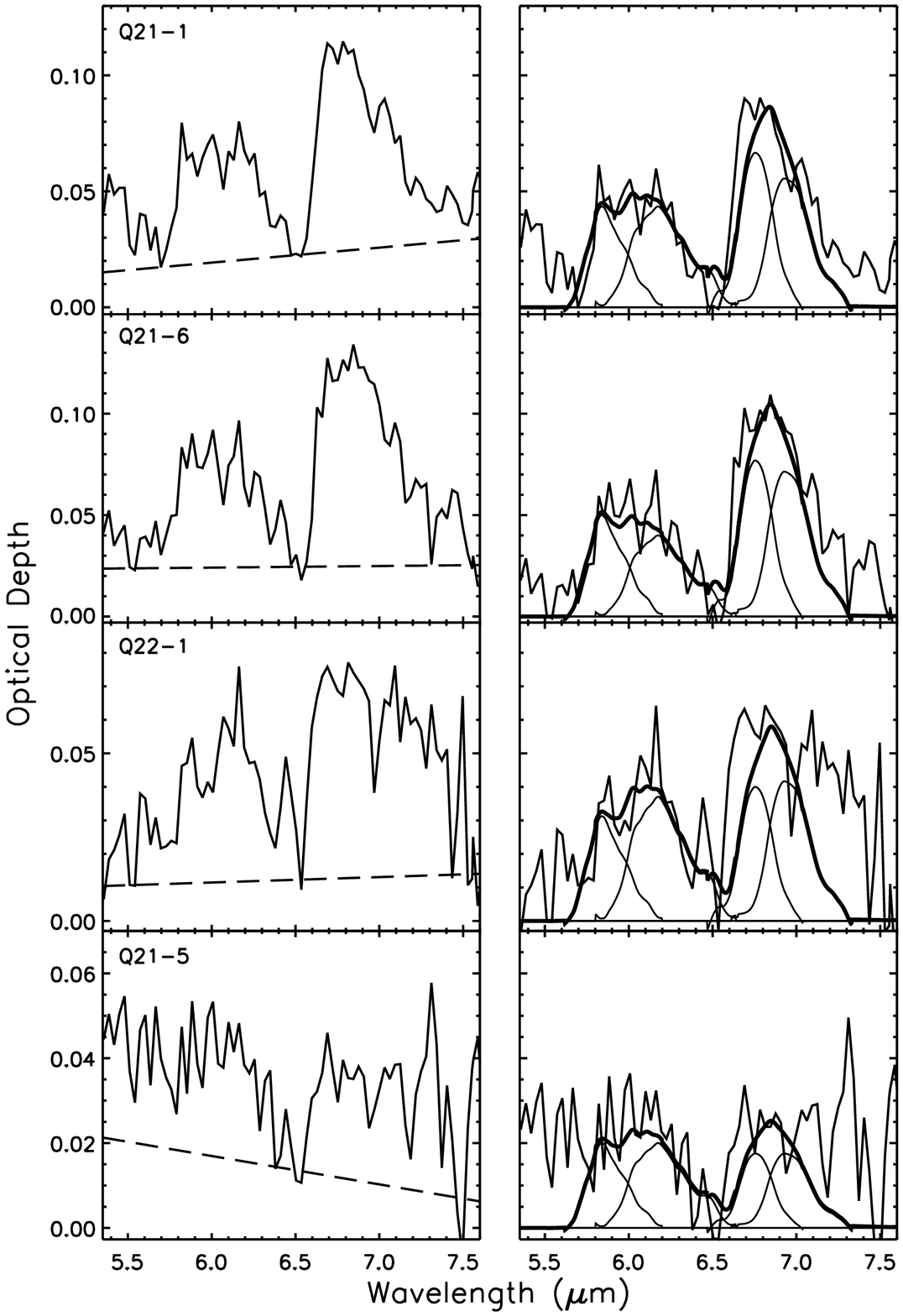}
\caption{Residual absorption in the 5.4-7.6 \micron\ region after subtraction of pure \water-ice.  [left] A local straight-line continuum (dashed line) is fitted across this spectral region (see text for details).  [right] Continuum-subtracted spectra. Thin lines represent components C1, C2, C3, C4 as described in the text.  The C5 component is not present in these sources.  Thick lines represent the sum of the four components.}
\label{fig:spectra58}
\end{figure}

It has been shown previously that the absorption feature at 6.0 \micron\ cannot be fully accounted for by \water-ice \nocite{tielens1984,keane2001,dhendecourt1996} (e.g., Tielens et al.\ 1984; Keane et al. 2001; d'Hendecourt et al.\ 1996).  In fact, if the 6.0 \micron\ absorption were fitted with \water-ice, the O-H stretching mode at 3.0 \micron\ is overestimated by up to a factor of 3 for high-mass young stellar objects \nocite{keane2001,schutte1996b,gibb2000}  (Schutte et al.\ 1996; Gibb et al.\ 2000; Keane et al.\ 2001).  While \methanol\ also has a strong band at 3.07 \micron\ (e.g., \nocite{hudgins1993} Hudgins et al.\ 1993) which can skew the apparent 3.0/6.0 ratio, this is unlikely to be a problem in quiescent lines of sight since the \methanol\ abundance, determined by analysis of absorption at 3.54 and 9.7 \micron, is less than 5\% of the \water-ice (see \S\ref{sec:traceices} and also Chiar et al.\ 1996).  Other proposed contributors to the absorption at 6.0 \micron\ are formic acid (HCOOH), formaldehyde (H$_2$CO), and the C-C stretch in aromatics.    Combining our small field star sample with the larger sample by Boogert et al. (2008; 41 low-mass YSOs, 8 high-mass YSOs and 2 field stars)\footnote{For sources with extremely deep 3.0 \micron\ bands ($\tau_{3.0}>5$), where the depths cannot be measured due to lack of detectable flux, Boogert et al.\ (2008) estimate $\tau_{3.0}$  using the H$_2$O libration mode at 13 \micron. This is the cause of the larger error bars on those data points.}, we examine the relationship between the 3.0 \micron\ \water-ice and the 6.0 \micron\ absorption bands.   Figure~\ref{fig:corr36} shows graphically, that the observed 6 \micron\ absorption cannot be fully accounted for by 10\,K \water-ice.     The shallower line shows the $\tau_{6.0}/\tau_{3.05}$ ratio for the 10\,K laboratory \water-ice analog from Hudgins et al.\ (1993).   The steeper line is the correlation line for the IC~5146 field stars.  Significantly, nearly all of the data points, including those of the field stars where little to no (energetic or thermal) processing of the ices is expected, lie above the laboratory ice line.   At larger values of $\tau_{3.05}$ (which roughly scales with \av), the scatter between the data points increases, indicative of increased alteration of the ices in the most heavily obscured YSOs.  

Boogert et al.\ (2008) recently decomposed the 6.0 and 6.85 \micron\ features into 5 components in addition to a contribution from \water-ice using spectra from a large sample of low-mass and high-mass YSOs.  They decompose the 6.0 \micron\ band into two components referred to as C1 ($\lambda=5.84$ \micron) and C2 ($\lambda=$ 6.18 \micron).  The C1 component is explained mostly by solid HCOOH and H$_2$CO with abundances up to 6\% relative to \water\ (Boogert et al.\ 2008).  The C2 component is also due to a blend of several species, with NH$_3$ being the most dominant.  Monomers, dimers and small multimers of \water\ mixed with CO$_2$ as well as contributions from salts are also possible (Boogert et al.\ 2008).   The 6.85 \micron\ absorption band shows great variation from source to source for the high-mass and low-mass YSOs that have been studied (Keane et al. 2001; Boogert et al. 2008).   This band is also decomposed into a short wavelength (C3, $\lambda=6.755$ \micron) and long wavelength (C4, $\lambda=6.943$ \micron) component by Boogert et al. (2008).  Both components are attributed to the ammonium ion, which would have to be produced by low-temperature reactions since both components are also detected in quiescent cloud spectra.   Underlying the C1 through C4 components is a fifth broad component (C5) whose origin may be related to processing of ices in the vicinity of the embedded YSOs (Boogert et al.\ 2008).

%% FIG 11: component correlations C1-C4, etc
\begin{figure}
\figurenum{11}
\includegraphics[angle=90,scale=0.35]{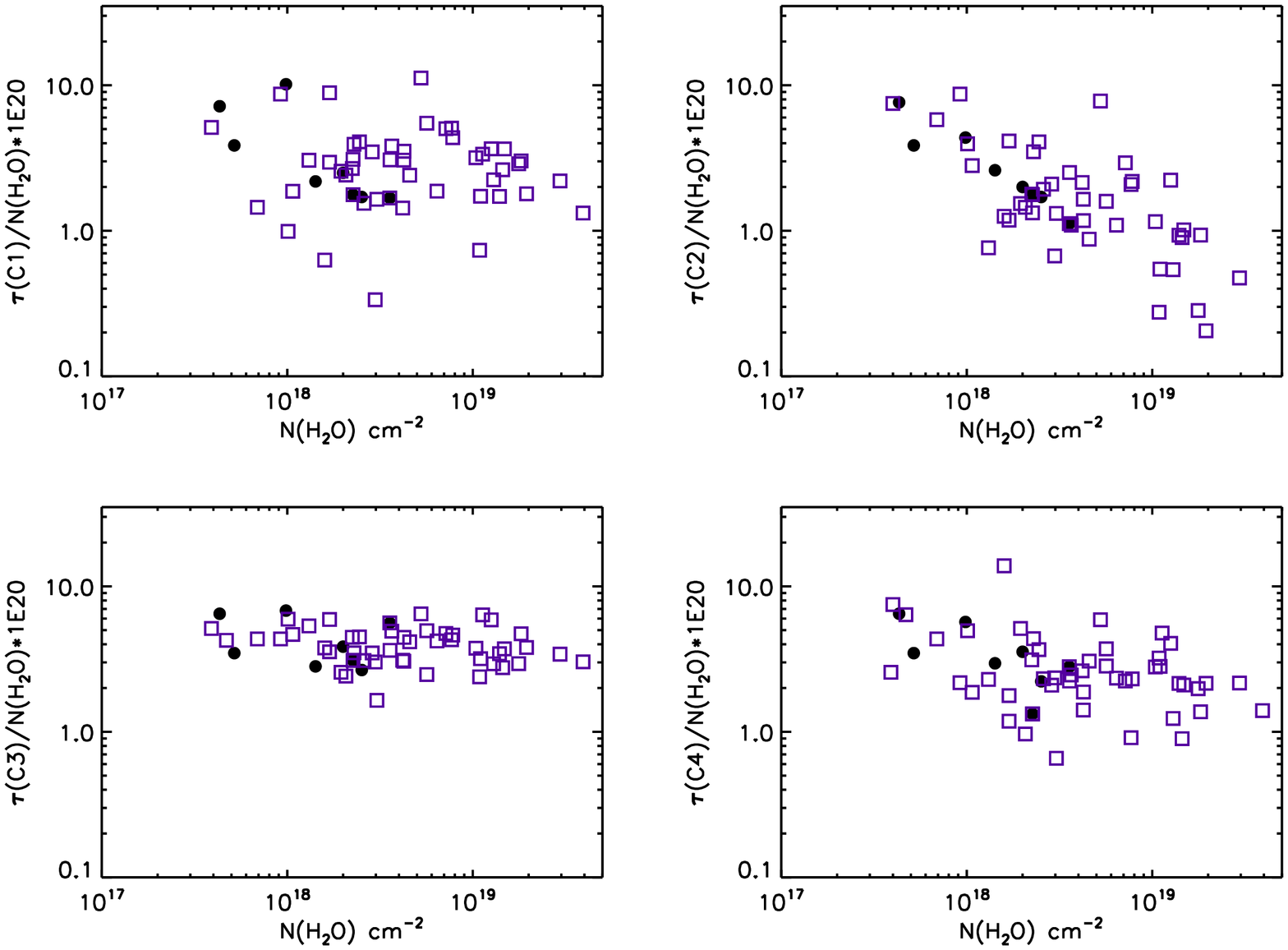}
\caption{The correlation of the C1, C2, C3 and C4 absorption components with \water-ice column density.  The optical depths of the C1 through C4 components are normalized to N(\water).  Purple square are the YSO data from Boogert et al.\ (2008) and solid circles are date for field stars from IC~5146 (this work), Elias 16 (Taurus, Boogert et al.\ 2008), EC118 (Serpens, Boogert et al.\ 2008). }
\label{fig:corr58}
\end{figure}

We use the component definitions as determined by Boogert et al.\ (2008) to investigate their presence in the field star spectra.  We show the component decomposition for the four sources with the highest S/N in Fig.~\ref{fig:spectra58}.  To be consistent with the determination of the C1 through C5 components in the studies by Boogert et al.\ (2008, 2011), we fit a local continuum across the 5 to 8 \micron\ region.   The shape of the 5 to 8 \micron\ region (where the absorption bands of photospheric \water\ and SiO appear) of the observed spectra is not always well-fitted by the computational models, so a straight-line continuum is fitted across this region.  This local continuum, fitted using points around 5.5, 6.5 and/or 7.5 \micron, removes the apparent offset of the optical depth spectra (see the left panels in Fig.~\ref{fig:spectra58}).   Following the subtraction of the local continuum, we decompose the 6.0 and 6.8 \micron\ features in their components (see the right panels in Fig.~\ref{fig:spectra58}). The C5 component is not apparent in any of our spectra, in line with the suggestion that this component is due to the presence of processed ices in the vicinity of embedded objects (Boogert et al.\ 2008).  We investigate the correlation of the C1 through C4 components with the \water-ice column density.  Along with the C1 through C4 measurements for the IC~5146 field stars in this paper, we also include the two field stars (Elias 16, EC118) and the YSOs from Boogert et al.\ (2008).    For consistency with the Boogert et al.\ studies (2008, 2011), we normalize the component depth (e.g., $\tau_{\rm C1}$) to N(\water) and plot that against N(\water) to trace the component growth relative to the ice column (Fig.~\ref{fig:corr58}).  The field stars (solid circles) seem to follow the trends previously noted for the YSOs.  We find that C2 shows a tight relationship of decreasing strength with increasing N(\water)  suggesting that this component is less volatile than \water-ice itself.   C1 and C4 show more scatter,. It is interesting to note that the C1 and C2 components seem to be of equal depth in each field star spectrum; this does not appear to be the case for the YSO spectra studied by Boogert et al.\ (2008) and is perhaps indicative of the sensitivity of one or both of these components to the UV or temperature processing that takes place in the environment around YSOs.  The (normalized) C3 component shows a flat relationship with  N(\water), linking it as a non-volatile ice component that it not sensitive to varying temperature or energetic conditions of YSO environments.

\subsection{Weak Ice Features in the 8.5 to 10 \micron\ region} 
\label{sec:traceices}
\begin{figure}
\figurenum{12}
\includegraphics[scale=0.4]{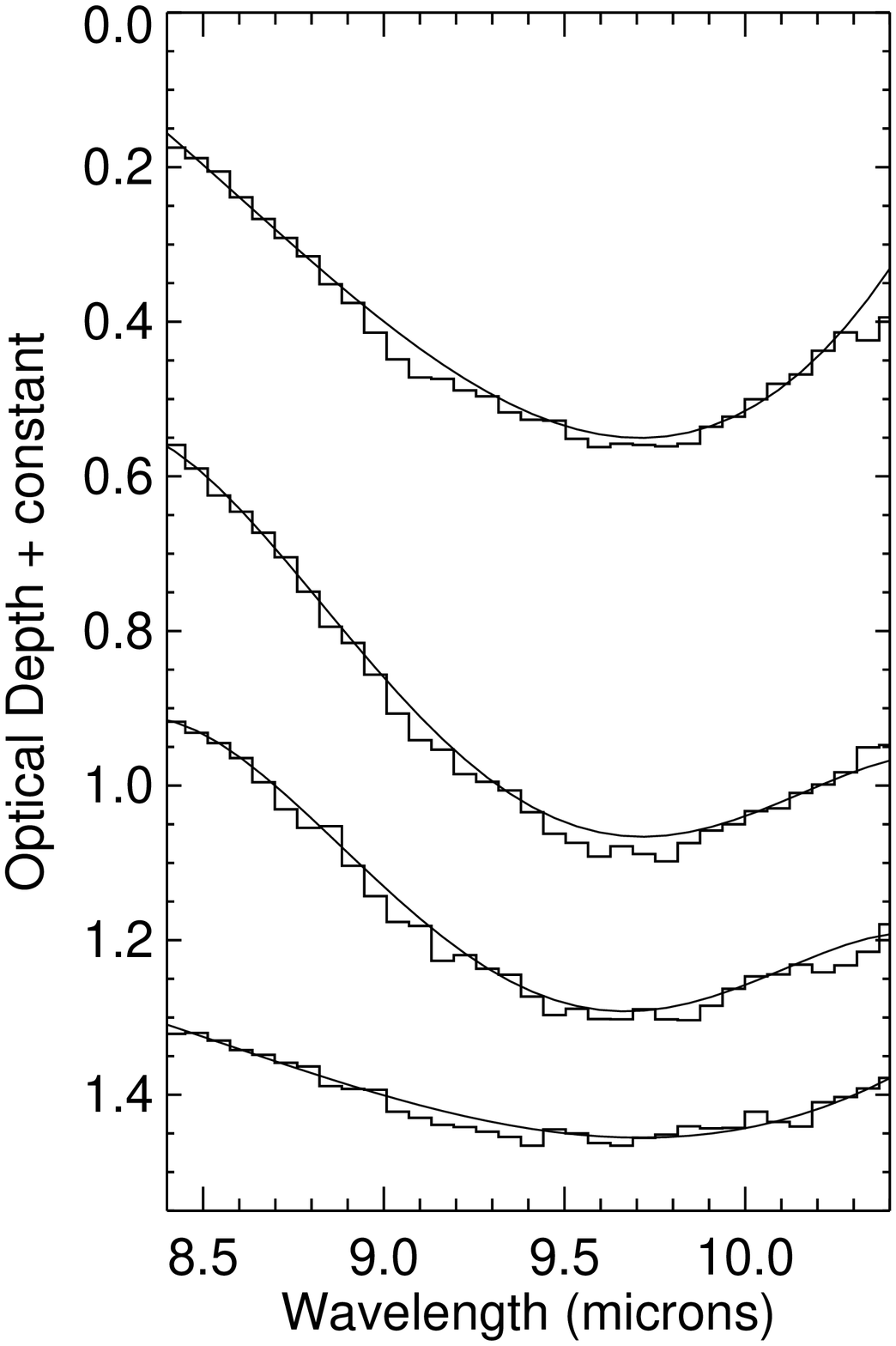}
\includegraphics[scale=0.4]{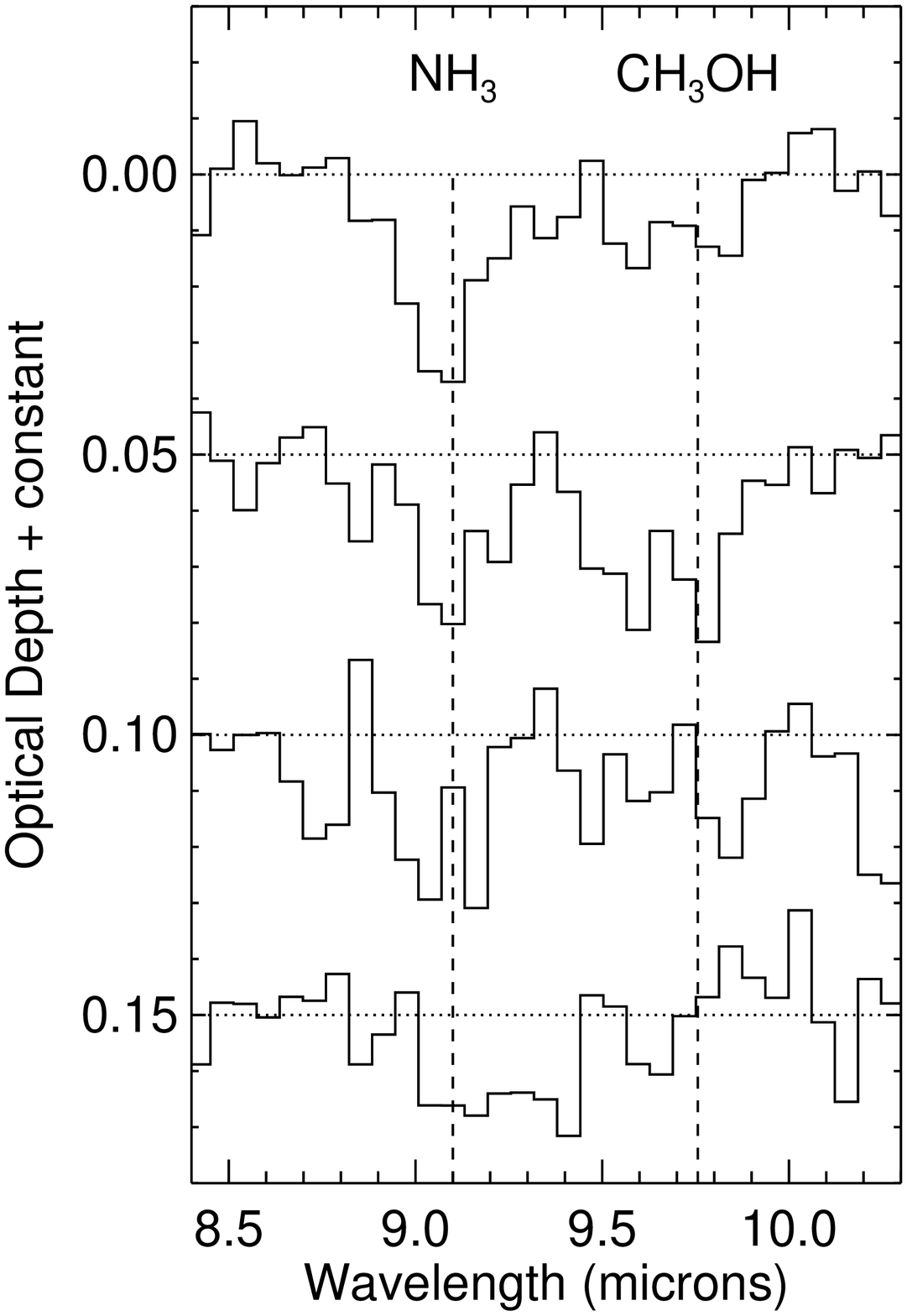}
\caption{[left] Silicate profiles for the sources (from top to bottom) Q21-1, Q21-6, Q22-1, and Q21-5.  Light line is the local polynomial continuum used to subtract off the silicate absorption to extract the weak ice features. [right] Continuum-subtracted optical depth spectra in the region of the silicate absorption. Sources are the same as in the left panel.   The approximate wavelength positions of NH$_3$ and CH$_3$OH are shown.  Note that the peak wavelength of NH$_3$ is extremely sensitive to temperature and neighboring molecules in the ice.}
\label{fig:silice}
\end{figure}

The 8.5 to 10 \micron\ region contains the  N-H vibration inversion mode of NH$_3$ \nocite{dhendecourt1986} (d'Hendecourt \& Allamandola 1986), the C-O stretching vibration of CH$_3$OH \nocite{sandford1993} (Sandford \& Allamandola 1993) as well as the O-O symmetric and asymmetric stretching mode of O$_3$ \nocite{bennett2005} (Bennett \& Kaiser 2005).    Substructure at these wavelengths appears at the bottom of the silicate feature for the most highly extincted sources, Q21-1, Q21-6 and possibly Q22-1.    We also examine the high  S/N spectrum of Q21-5.  For this source, which has relatively lower extinction  (see Table~\ref{table:sources}), we do not expect to see the weak ice absorptions in the 8.5 to 10 \micron\ region.  So, we use this source as a check that the substructure detected in the high extinction sources is not due to an artifact in the data reduction process.  For all four sources, we fitted and subtracted a local polynomial continuum in the 8.4-10.5 \micron\ region of the optical depth spectra in order to subtract the contribution of the broad silicate absorption. The continuum fits and resulting residuals are compared in Fig.~\ref{fig:silice} in the left and right panels, respectively.  In the spectra of Q21-1, Q21-6, Q22-1, an absorption feature appears at 9.0 \micron\ with FWHM of about 24 cm$^{-1}$.  A second absorption feature at 9.7 \micron, with FWHM of $\sim38$ \invcm\ is also present (tentatively in Q 22-1).  We discuss the likely identifications of there features below.

%% FIG 12: 9 mic ices

We attribute the weak absorption at $\sim9.7$ \micron, present in the spectra of Q 21-1 and Q 21-6 (and tentatively Q22-1), to the C-O stretching vibration in \methanol-ice, that occurs at about 9.75 \micron\ (Sandford \& Allamandola 1993).  As shown in Fig.~\ref{fig:icewing}, the \methanol\ identification is supported by residual absorption in the long-wavelength wing of the 3 \micron\ \water-ice feature that is not accounted for by the typical 3.47 \micron\ ``hydrocarbon ice'' feature.  Assuming a FWHM equal to 29 \invcm\ and intrinsic strength of $A=1.8\times10^{-17}$ cm molecule$^{-1}$ (d'Hendecourt \& Allamandola 1986), we calculate the column density of \methanol\  for the three lines of sight (Table~\ref{table:cds}).  The \methanol/\water\ ratio is in the range of 1-2 \% for these sources.  This is well within the previous limiting value ($<5\%$)  determined for the Taurus dark cloud \nocite{chiar1996} (Chiar et al.\ 1996).

The N-H inversion mode occurs near 9.1 \micron, and its central wavelength is highly sensitive to matrix neighbors.  Embedded in a polar matrix, the feature peaks at 8.97 \micron, and in an apolar matrix, the feature peaks at 9.57 \micron\ \nocite{lacy1998}(Lacy et al. 1998).   The observed 9.0 \micron\ feature falls in the range of the N-H umbrella mode, although it is narrower than the available NH$_3$ ice analogs which have widths equal to $\sim 50$ cm$^{-1}$ (d'Hendecourt \& Allamandola 1986; Lacy et al. 1998).  Given the difficulty in extracting the ice feature from the bottom of the broad silicate absorption, it is possible that we have underestimated the width by a factor of 2.  Taking this uncertainty into account and using an intrinsic strength of $A=1.7\times10^{-17}$ cm molecule$^{-1}$ for the $\nu_2$ umbrella mode of NH$_3$ (d'Hendecourt \& Allamandola 1986), we compute column densities for Q 21-1, Q 21-6 and Q 22-1.  These are given in Table~\ref{table:cds}.  For all three sources, the column of NH$_3$ relative to \water-ice ranges from 2 to 5\%.  

Ozone (\ozone) ice has absorption features at 9.06 \micron\ and 9.64 \micron\ due to the O-O symmetric and asymmetric stretch \nocite{bennett2005} (Bennett \& Kaiser 2005).   The O-O asymmetric stretch is $\sim80$ times stronger than that of the corresponding symmetric stretch (Bennett \& Kaiser 2005).  Given the depth of the observed absorption centered at 9.7 \micron, \ozone-ice would contribute a negligible amount to the optical depth at 9.0 \micron.    Thus, it is unlikely that \ozone\ contributes significantly to the ices in these lines of sight.

\section{Extinction}
\label{sec:extinction}

In the diffuse ISM, the extinction due to silicates (i.e., $\tau_{9.7}$) shows a tight linear correlation with \av\ (Whittet 2003 and references therein).  Whittet et al. (1988) and Chiar et al.\ (2008) showed that, in dense clouds, the optical depth of the 9.7 \micron\ silicate feature ($\tau_{\rm silicate}$) is weaker per unit \av\ compared to the diffuse ISM.   The Chiar et al. (2007) study analyzed 30 spectra toward six different dense clouds, including IC 5146.  For that sample, $\tau_{9.7}$ was computed using a low-order polynomial fitted over the 5 to 15 \micron\ region and an estimate of \ejk\ based on a likely range of spectral types for the background stars.  The $\tau_{9.7}$ and \ejk\ estimates thus vary slightly from the more thorough analysis presented here; most values are within 20\% of those determined previously.    A recent reanalysis of the Chiar et al.\ (2007) sample reports that  the {\em total} optical depth at 9.7 \micron\  ($\tau_{9.7}$) correlates with the extinction in the K-band, $A_K$ \nocite{mcclure2009} (McClure 2009), and the data points lie along the same correlation line as for the diffuse ISM.  However, there are some caveats to this illusory agreement in correlations.  First,  the total $\tau$ at 9.7 \micron\ includes the underlying ``continuum'' extinction, the extinction due to the silicates themselves, and a small contribution from ices (for the high extinction sources) so a correlation between $\tau_{9.7}$ and $A_K$ does not speak directly to the relation between the silicate and K-band extinction. Second,  for the diffuse ISM studies, the optical depth due to the silicates alone (i.e., $\tau_{\rm silicate}$) is shown to correlate with the visual extinction (Roche \& Aitken 1984; Rieke \& Lebofsky 1985; Bowey et al.\ 2004; Chiar \& Tielens 2006).  In the diffuse ISM studies, continuum extinction due to the ISM dust and the dust shells of the Wolf-Rayet stars is accounted for by a power-law extinction curve or $1/\lambda$ emissivity curve, respectively.  

\begin{figure}
\figurenum{13}
\includegraphics[angle=0,scale=0.45]{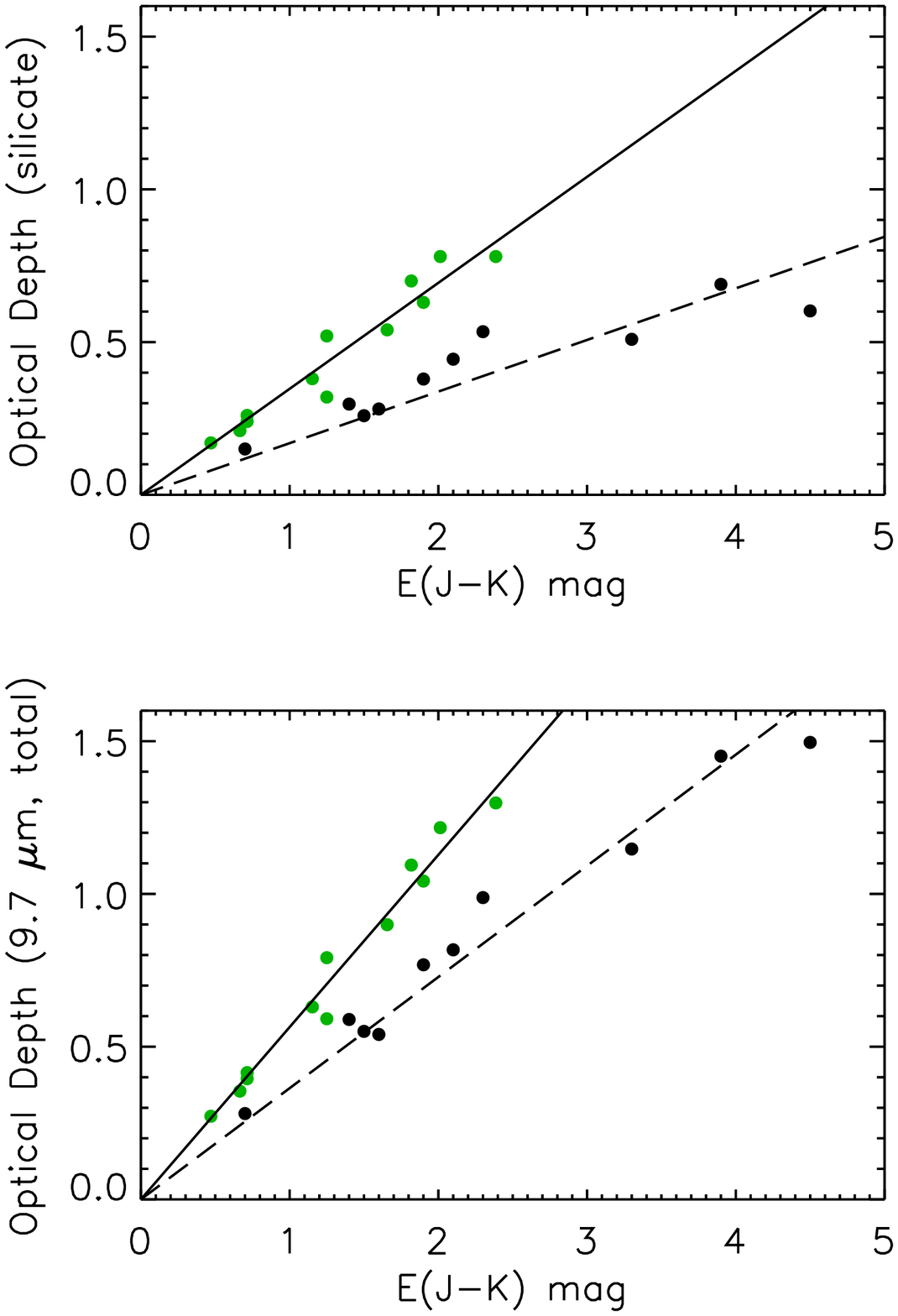}
\caption{[top] The optical depth of the silicate feature at 9.7 \micron\ versus the \ejk\ color excess.  [bottom] The total optical depth at 9.7 \micron\ (including continuum extinction) vs \ejk.  Green points are for the diffuse ISM  (Whittet 2003 and references therein).  In the bottom frames, the diffuse ISM  $\tau$ values (green circles) do not include continuum extinction (see text for details).}
\label{fig:sil_extinction}
\end{figure}

We illustrate the relationship between the extinction quantities in Fig.~\ref{fig:sil_extinction}.   Optical depth is related to the extinction by $\tau_{\lambda} = -\ln (10^{-A_{\lambda}/2.5})$.  The upper frame in Fig.~\ref{fig:sil_extinction}  plots $\tau_{9.7}$,  the peak silicate optical depth excluding the continuum extinction, versus the \ejk\ color excess for the IC~5146 sources and diffuse ISM sources (from Whittet 2003 and references therein).  As discussed previously by Chiar et al.\ (2007),   the diffuse ISM sources follow a tighter and steeper correlation compared to the IC 5146 sources.  In the lower panel of Fig.~\ref{fig:sil_extinction}, we follow McClure's method and plot the {\em total} optical depth at 9.7 \micron\ (including continuum and silicate extinction) for the IC 5146 sources and the diffuse ISM sources.  The continuum extinction at 9.7 \micron\ is calculated using the Indebetouw et al.\ (2005) extinction law ($A_{\lambda}/A_K=0.43$ at 8 \micron), assuming $A_{9.7} = A_{8}$, where $A_K$ is determined from the continuum fits shown in Fig.~\ref{fig:continua}. The correlation lines are least-squares fits to the data and show that including continuum extinction shifts both the diffuse ISM and dense cloud points up in the $\tau_{9.7}$ vs. $E(J-K)$ plot.

%% FIGURE 13: correlation silicate vs e(j-k)

%%%% TABLE 3
\begin{deluxetable}{ccccc}
\tabletypesize{\scriptsize}
%\rotate
\tablecaption{Column Densities of Various Ice and Dust Components\label{table:cds}}
\tablewidth{0pt}
\tablehead{
\colhead{Source ID} & \colhead{N(H$_2$O)\tablenotemark{a}} & \colhead{N(CO)}  & \colhead{N(NH$_3$)} 
                                     & \colhead{N(CH$_3$OH)} \\
\colhead{} & \multicolumn{4}{c}{$\times 10^{17}$ cm$^{-2}$} 
}
\startdata
Q21-1  & 25.2  & 6.83 & 0.6 - 1.1 & 0.24  \\
Q21-2  &  9.84 & \nodata &  &  \\
Q21-3  &  \nodata & \nodata &  &   \\
Q21-4  & 5.18 & \nodata   & &  \\
Q21-5  &  5.18 & \nodata  & &  \\
Q21-6  & 20.0  & \nodata & 0.4 - 0.8 & 0.32  \\
Q22-1  & 14.2 & \nodata  & 0.3 - 0.7 & 0.16:  \\
Q22-3  & \nodata & \nodata &  &  \\
Q23-1  & 4.32 & \nodata  & &  \\
Q23-2  &   \nodata & \nodata & &  \\
\enddata
\tablenotetext{a}{No SpeX data exist for Q21-3, Q22-3, Q23-2.}
\tablenotetext{b}{colon denotes uncertain value}
\end{deluxetable}

%%%% TABLE 4
\begin{deluxetable*}{lcccc}
\tabletypesize{\scriptsize}
\tablecaption{Relative Ice Abundances in Different Environments}
\tablewidth{0pt}
\tablehead{
\colhead{Ice} & \colhead{IC 5146} & \colhead{Taurus} & \colhead{low-mass YSOs} & \colhead{high-mass YSOs}
}
\startdata
H$_2$O        & 1                   & 1                &   1       & 1 \\
CO                 & 0.27              &  $0.32\pm0.12$         &  0.32  &   $0.12\pm0.07$  \\
CO$_2$        & 0.35              &  0.19         &  $0.19\pm0.13$          &  $0.18\pm0.07$  \\
CH$_3$OH  & $\sim0.02$  & $<0.05$   &  0.08   & $<0.04-0.32$ \\
NH$_3$        & $\sim0.05$  & $<0.10$   &   $<0.27$          & $<0.02-0.17$ \\
OCN$^-$      & $<0.01$       & $<0.01$   & $<0.002-0.057$  & $<0.002-0.063$  \\
\enddata
\end{deluxetable*}

\section{Summary and Discussion}
\label{sec:final}

Most comparisons of the relative abundances, temperature, and evolution of ices in star-formation regions are made with respect to the Taurus dark cloud which is used throughout the literature as the prototypical pristine environment for the study of dust and ices.  This is because the fortuitous placement of bright field giants behind the Taurus cloud have made spectroscopic studies practical.     In this paper, we present a spectroscopic study of the quiescent IC 5146 dark cloud which was made possible by the sensitive IRS instrument on Spitzer and the SpeX instrument on the IRTF.   We have shown that the quiescent IC 5146 dark cloud shares many of the characteristics of the Taurus cloud; we summarize our results here.    

IRTF-SpeX  2 \micron\ spectra were used to classify the IC 5146 background stars with stunning accuracy, thereby allowing continua to be fitted over the full 2 to 20 \micron\ range, resulting in reliable determinations  of the extinction,  ice and silicate absorption profiles.  We show that, after taking foreground extinction into account, the \water-ice extinction threshold in the IC 5146 dark cloud is nearly equivalent to that in Taurus, showing the universality of the onset of ice mantle growth in regions far from internal sources of radiation.  The growth of  the icy mantles as measured by the increase of  $\tau_{3.0}$ with respect to \av\ is similar in both clouds as shown by the comparable slopes of the regression lines on a $\tau_{3.0}$--\av\ plot.

We have compared ice abundances in the IC 5146 cloud to Taurus and low- and high-mass YSOs (Table~4).  Previous estimates of the abundances of \methanol\ and \ammonia\ in dense clouds were based on limits only \nocite{chiar1996,gibb2001} (Chiar et al.\ 1996; Gibb et al.\ 2001).  However, recent laboratory \nocite{bottinelli2010} (Bottinelli et al.\ 2010) and computational simulations \nocite{cuppen2009} (Cuppen et al.\ 2009) suggest that the hydrogenation process that forms \methanol\ should be efficient in cold environments.   The recent observational study by Boogert et al.\ (2011) of ices in isolated dense cores finds \methanol\ abundances in the 5 to 12\% range, in line with the idea of efficient production of \methanol\ in the coldest environments.  Here we have shown that \methanol\ and \ammonia\ are present in the quiescent cloud medium at the 2 to 5\% level relative to \water-ice.   In general, \methanol\ ice is more abundant in YSO environments compared to quiescent dark clouds,  The \ammonia\ abundances for the dense cloud and YSO environments shows wide variation, perhaps reflecting the difficulty of the spectroscopic measurement rather than the sensitivity of \ammonia-ice itself to cloud conditions.  In either case, \ammonia-ice is an important mantle constituent in all dense cloud environments, and therefore an important reservoir of elemental N in these regions.  The CO-ice abundance appears to be similar in the IC 5146 and Taurus clouds, while the CO$_2$ abundance is elevated in IC 5146 compared to Taurus.  Whittet et al.\ (2009) attribute this difference to variations in gas-phase chemistry, but point out that the amount of oxygen depleted into the ices is consistent in the two clouds.   As has been previously noted \nocite{kerr1993,thi2006} (Kerr et al.\ 1993; Thi et al.\ 2006), the YSOs show a large variation in CO abundance, reflecting the sensitivity of CO to the environmental conditions (temperature, UV), while the CO$_2$ abundance is remarkably consistent (with the exception of the IC 5146 measurement!).   

Finally, our results firmly support the conclusions by Chiar et al.\ (2007) that the silicate optical depth, $\tau_{9.7}$, grows more slowly per unit \av\ in dense clouds relative to the diffuse ISM.   The change in dust characteristics that cause the difference in the $\tau_{9.7}$--\av\ relation is still an open question.

\acknowledgments   This work is based [in part] on observations made with the Spitzer Space Telescope, which is operated by the Jet Propulsion Laboratory, California Institute of Technology under a contract with NASA. Support for this work was provided to JEC by NASA through an award issued by JPL/Caltech and to YJP, LJA, KE, TPG, TLR, SAS by NASA.  LJA gratefully acknowledges support from NASA's Astrobiology (Grant 811073.02.12.03) and Laboratory Astrophysics (Grant 09-APRA09-0019) programs.  DCBW is grateful to the NASA Exobiology and Evolutionary Biology Program (grant NNX07AK38G) and the NASA Astrobiology Institute (grant NNA09DA80A) for financial support.  REM and TRG are supported by the Gemini Observatory, which is operated by the Association of Universities 
for Research in Astronomy, Inc., on behalf of the international Gemini partnership of Argentina, 
Australia, Brazil, Canada, Chile, the United Kingdom, and the United States of America.    The work is also based in part on observations made at the Infrared Telescope Facility, which is operated by the University of Hawaii under Cooperative Agreement no. NCC 5-538 with the National Aeronautics and Space Administration, Science Mission Directorate, Planetary Astronomy Program.  This publication makes use of data products from the Two Micron All Sky Survey, which is a joint project of the University of Massachusetts and the Infrared Processing and Analysis Center/California Institute of Technology, funded by the National Aeronautics and Space Administration and the National Science Foundation.


%% in parentheses, after the keyword, but they will not be verified.

{\it Facilities:} \facility{Spitzer (IRS)}, \facility{IRTF (SpeX)}



\nocite{gerakines1995,gerakines1995}

%\bibliographystyle{apj}
%\bibliography{references}

\end{document}